\documentclass[11pt]{article}

\usepackage{fullpage}
\usepackage{float}
\usepackage[font=small]{caption}

\usepackage{latexsym}
\usepackage{amssymb}
\usepackage{amsmath}
\usepackage{enumerate}
\usepackage{mathtools}
\usepackage{verbatim}
\usepackage{braket}
\usepackage{xcolor}
\usepackage{url}	
\usepackage{graphicx} 
\usepackage{tikz-cd}
\usepackage[colorlinks=true, linkcolor=blue]{hyperref}

\def\X{\mbox{$\cal X$}}

\def\F{\Bbb F}
\def\CN{\Bbb C}

\def\F{\Bbb F}

\newtheorem{theorem}{Theorem}[section]
\newtheorem{lemma}[theorem]{Lemma}
\newtheorem{definition}[theorem]{Definition}
\newtheorem{corollary}[theorem]{Corollary}

\newtheorem{openproblem}[theorem]{Open Question}
\newtheorem{algorithm}[theorem]{Algorithm}
\newtheorem{fact}[theorem]{Fact}

\def\X{\chi}

\def\finito{{\hspace*{\fill}  \mbox{$\blacksquare $}}}

\begin{document}
\author{Louay Bazzi
\footnote{
Department of Electrical and Computer Engineering, American University of Beirut, 
Beirut, Lebanon. E-mail: lb13@aub.edu.lb.} 
}
\title{
Improved decoding algorithms for surface codes under independent bit-flip and phase-flip errors
}


\maketitle
\begin{abstract}
We study exact decoding for the toric code and for planar and rotated surface codes under the
standard independent \(X/Z\) noise model, focusing on Separate Minimum Weight  (SMW) decoding and
Separate Most Likely Coset (SMLC) decoding.
For SMW decoding, we show that an \(O(n^{3/2}\log n)\)-time decoder is achievable, improving over the \(O(n^{3}\log n)\) worst-case time of the standard 
approach based on complete decoding graphs.
Our approach is based on a local reduction of SMW decoding to the minimum weight perfect matching
problem using Fisher gadgets, which preserves planarity for planar and rotated surface codes and
genus~\(1\) for the toric code.
This reduction enables the use of Lipton--Tarjan planar separator method and implies that SMW
decoding lies in \(\mathrm{NC}\). 
For SMLC decoding, we show that the planar surface code admits an exact decoder with
\(O(n^{3/2})\) algebraic complexity and that the problem lies in \(\mathrm{NC}\),
improving over the \(O(n^{2})\) algebraic complexity of Bravyi \emph{et al.}
Our approach proceeds via a dual-cycle formulation of coset probabilities and an explicit reduction
to planar Pfaffian evaluation using Fisher--Kasteleyn--Temperley construction.
The same complexity measures apply to SMLC decoding of the rotated surface code.
For the toric code, we obtain an exact polynomial-time SMLC decoder with
\(O(n^{3})\) algebraic complexity. 
While the SMLC formulation is motivated by connections to statistical mechanics, we
provide a purely algebraic derivation of the underlying duality based on MacWilliams duality and
Fourier analysis.
Finally, we discuss extensions of the  framework to the depolarizing channel.

\end{abstract}

\tableofcontents

\section{Introduction}

Quantum error correction  is  essential for scalable fault-tolerant quantum computation.
The  theory of quantum error-correcting codes 
was initiated by 
Shor’s discovery that quantum information can be protected against noise using redundancy~\cite{Shor1995,Shor1996}.
 The subsequent Calderbank--Shor CSS construction and Steane’s independent work~\cite{CalderbankShor1996,Steane1996}
established a  bridge from classical linear codes to quantum codes, while Gottesman’s stabilizer formalism~\cite{Gottesman1997}
provided an  algebraic framework for  broad families of quantum codes.

Soon thereafter, Kitaev introduced the toric code~\cite{Kitaev1997}, a CSS  code that initiated the two-dimensional topological paradigm for quantum error correction. 
Planar realizations of the toric code, commonly referred to as surface codes~\cite{BravyiKitaev1998,Fowler2012SurfaceCodes},
are compatible with local qubit architectures and have become leading candidates for practical
fault-tolerant quantum computation. 
Among these, the planar surface code 
\cite{BravyiKitaev1998, Dennis2002} 
and its rotated variant 
\cite{Bombin2007,rotated2} 
provide the most basic planar realizations of surface codes and serve as canonical examples throughout the literature. 
A key insight in this line of work is the connection between decoding and statistical mechanics:
Dennis \emph{et al.}~\cite{Dennis2002} related toric/surface-code decoding thresholds to an Ising-model formulation
along the Nishimori line.

 Decoding is the  algorithmic task of inferring a recovery consistent with a measured syndrome according to a specified objective.
 Throughout this work, we assume perfect syndrome measurements and focus on decoding with noiseless
syndrome information. 
We mostly work in the standard independent \(X/Z\) noise model for CSS codes, where \(X\)- and \(Z\)-errors are generated according to independent binary symmetric channels, so the decoding problem separates into independent \(X\)- and \(Z\)-subproblems. 
Within this setting we focus on two objectives:
\emph{Separate Minimum Weight (SMW)} decoding, which seeks a 
Pauli error of minimum weight consistent with the syndrome,
and \emph{Separate Most Likely Coset (SMLC)} decoding, which seeks the most likely stabilizer coset consistent with
the syndrome. 
In the independent \(X/Z\) noise setting for CSS codes, these objectives are decorrelated versions of the \emph{Minimum Weight (MW)} and \emph{Most Likely Coset (MLC)} decoding problems.
The latter problems correspond to the depolarizing noise model, in which qubits undergo \(X\)-, \(Y\)-, and \(Z\)-errors with equal probability, and are discussed later in the paper.

Many surface code decoders have been proposed; see~\cite{deMartiOlius2024} for a survey.
These include \emph{Minimum Weight Perfect Matching (MWPM)} decoders,
fast approximate decoders such as Union-Find~\cite{DelfosseNickerson2021},
tensor-network-based approach~\cite{MLCSurface},
and other decoding strategies.  
The present work is concerned with \emph{exact} algorithms for SMW and SMLC decoding,
and with improving their asymptotic runtime by exploiting the planar or low-genus structure
of surface and toric codes.

For SMW decoding, the standard  approach is to reduce it to an instance of the  MWPM problem by constructing a complete graph on the syndrome defects, 
known as the \emph{decoding graph}, with edge weights given by all-pairs shortest-path distances,
and then applying a general-purpose blossom  algorithm to solve the resulting  matching problem.
The worst-case complexity of blossom-based MWPM decoding is \(O(n^{3}\log n)\)~\cite{Higgott2022,HiggottGidney2023}, where \(n\) is the number of qubits,
and several improved implementations with stronger empirical performance have been developed
\cite{Fowler2015,Higgott2022,WuLiyanageZhong2022,HiggottGidney2023,WuZhong2023}.

For SMLC decoding of the planar surface code, Bravyi \emph{et al.}~\cite{MLCSurface} gave an exact
polynomial-time algorithm by reducing coset probabilities to determinant and Pfaffian evaluations
of structured matrices. The algebraic complexity of their algorithm is \(O(n^{2})\) arithmetic
operations. Accounting for the cost of performing arithmetic operations exactly, the resulting
bit complexity is \(\tilde{O}(n^{3})\), where \(\tilde{O}(f(n)) = O\!\left(f(n)\log^{O(1)} n\right)\).

This paper revisits both objectives from a unified graph-theoretic perspective, with an additional
algebraic component for the SMLC problem.
The resulting contributions are summarized below, starting with asymptotic improvements.

\subsection{Overview of results and techniques}

In terms of asymptotic improvements, we reduce the SMW decoding time to
\(O(n^{3/2}\log n)\) for the toric code and for planar and rotated surface codes,
and the SMLC decoding time for the planar surface code to
\(O(n^{3/2})\) arithmetic operations, 
 which corresponds to
\(\tilde{O}(n^{3/2+1})\) bit operations.
We further show that the SMLC decoding problem for the rotated surface code can be solved within the
same asymptotic complexity, and that for the toric code it can be solved in
\(O(n^{3})\) arithmetic operations.
We also conclude that SMW and SMLC decoding for planar and rotated surface codes lies in
\(\mathrm{NC}\).
Consequently, these decoding problems are highly parallelizable, in the sense that they admit
polylogarithmic-depth circuits.
We summarize below the underlying techniques and constructions.

A key observation is that the standard reduction of SMW decoding to MWPM, via the construction of a
complete decoding graph on syndrome defects, does not retain  the local geometry and the planar or low-genus
structure of the underlying code lattice.
For SMW decoding, we instead reduce the problem to MWPM on sparse \emph{decorated graphs} associated
with the primal and dual lattices.
These graphs are constructed by replacing each vertex of the lattice with a constant-size
\emph{Fisher gadget}~\cite{Fisher66} that encodes the local parity constraints imposed by the syndrome
and, in planar settings, the boundary conditions of the code.
Unlike the standard approach based on complete decoding graphs, the decorated graphs preserve the
local geometry and structure of the lattice: they are planar for planar and rotated surface codes and have
genus~\(1\) for the toric code.

This structural property enables the use of the Lipton--Tarjan \emph{planar separator} method~\cite{LiptonTarjan80}
to achieve improved decoding runtimes.
As a result, SMW decoding can be performed in \(O(n^{3/2}\log n)\) time for the toric code
(Theorem~\ref{toricSMW}) and for planar and rotated surface codes (Theorem~\ref{planarSMW}),
improving on the \(O(n^{3}\log n)\) worst-case complexity of blossom-based MWPM on complete graphs.

For planar and rotated surface codes, we further conclude that SMW decoding lies in \(\mathrm{NC}\),
by invoking the breakthrough result of Anari and Vazirani~\cite{AV18} that MWPM on planar
graphs lies in \(\mathrm{NC}\) (Theorem~\ref{planarSMW}).

We optimize the decorated-graph construction by first allowing limited local nonplanarity without
sacrificing the applicability of separator-based methods.
We further reduce the size of the construction by showing that there exists a minimum weight
\(2\)-chain satisfying the syndrome with exactly one edge incident to each syndrome defect, rather
than merely satisfying the odd-degree constraint. 
This property is established via structural \emph{degree-\(3\)-defect resolution} theorems for the toric
code (Theorem~\ref{DJoin}) and for the planar and rotated surface codes
(Theorems~\ref{DJoinPlanar} and~\ref{DJoinRotated}), which are of independent interest. 
For the toric code, the corresponding result holds under the assumption that the lattice length 
is even.

Our SMLC decoder for planar and rotated surface codes is rooted in the statistical-physics
perspective on surface-code decoding, where coset probabilities have an interpretation in terms of
Ising-type models on the Nishimori line, as identified by Dennis \emph{et al.}~\cite{Dennis2002}.
The resulting representation admits the classical Kramers--Wannier high-temperature graphical
expansion into weighted even subgraphs~\cite{KramersWannier1941,KramersWannier1941PartII}, which leads
to an expression of coset probabilities as weighted sums over all relative dual cycles.
While we were originally led to this 
\emph{dual-cycle expansion}
 via statistical-physics considerations, we establish it
in Lemma~\ref{srufaceSMLCDual} through a more general statement (Lemma~\ref{genlem}) with a simpler
proof based on MacWilliams duality and Fourier analysis.
The proof itself is  independent of Ising-model or Nishimori-line arguments and applies to
arbitrary linear codes.
In this proof route, the geometric interpretation specific to surface codes enters only through the
CSS definition of these codes.

Once coset probabilities are expressed in terms of dual cycles, we reduce the problem to a 
\emph{Pfaffian}
formulation by converting the dual-cycle expansion into a weighted sum over perfect matchings on a
\emph{planar decorated} graph associated with the dual lattice, using Fisher gadgets in their original role
~\cite{Fisher66}.
Conceptually, the decorated graphs used here are closely related to those introduced for
SMW decoding, differing primarily in the weighting scheme and in the quantities computed on them.
This enables the use of the classical Fisher--Kasteleyn--Temperley framework~\cite{Kasteleyn1963},
whereby the matching sum is computed via a Pfaffian.

To evaluate the Pfaffian, we employ Yuster’s algorithm~\cite{Yuster} for computing the absolute value of the determinant of a matrix supported by a planar graph.
This algorithm is based on the seminal nested dissection method of Lipton, Rose, and Tarjan~\cite{LiptonRoseTarjan79}.
Using this approach, we obtain in Theorem~\ref{planarRotSMLC} an exact SMLC decoder for planar and rotated surface codes with
\(O(n^{\omega/2})\) algebraic complexity and \(\tilde{O}(n^{\omega/2+1})\) bit complexity,
where \(\omega\) is the matrix multiplication exponent.

We also conclude that the problem lies in \(\mathrm{NC}\), since determinant computation is 
 in \(\mathrm{NC}\)~\cite{Csa76}.

Algorithmically, a key advantage of this formulation is that the planar Pfaffian structure is made
explicit, enabling the use of determinant and Pfaffian algorithms for matrices with planar support,
thereby leading to an improvement in decoder complexity by a factor of 
\(\Theta(\sqrt{n})\) 
over the approach
of Bravyi \emph{et al.}~\cite{MLCSurface}.
(While the algebraic complexity reported in~\cite{MLCSurface} is \(O(n^{2})\), it  reduces to
\(O(n^{(\omega+1)/2})\) when fast matrix multiplication algorithms are used.)

We then extend the SMLC framework to the toric code, obtaining an exact polynomial-time decoder with
\(O(n^{3})\) algebraic complexity and \(\tilde{O}(n^{4})\) bit complexity
(Theorem~\ref{troicSMLC4}).
To our  knowledge, 
no exact polynomial-time SMLC decoder for the toric code appears to have been previously known. 
The increased complexity arises 
because the corresponding reduction is no longer planar but instead lives on a surface of genus~\(1\), so the weighted sum over perfect matchings is no longer given by
a single Pfaffian.
Instead, following the extension of Kasteleyn’s theory to bounded-genus graphs by Galluccio and
Loebl~\cite{GalluccioLoebl}, it can be expressed as a linear combination of four Pfaffians.
In this setting, resolving Pfaffian signs is more costly than computing absolute values, leading to
a higher overall complexity.

The results reported in this paper were originally motivated by the study of MW decoding under the
\emph{depolarizing channel} model.
Joint decoding under depolarizing noise requires reasoning
locally about the coupling between \(X\)- and \(Z\)-errors on each qubit, a locality that is lost 
in classical reductions based on complete decoding graphs.
This issue led us to a local, geometry-preserving perspective based on decorated graphs and
small gadgets.
While this local framework does not presently yield an efficient solution to MW  decoding
under depolarizing noise, it proved instrumental in obtaining the improvements for the independent
noise model reported in this paper.
Within this framework, we formulate both the MW and MLC decoding problems for surface codes under
depolarizing noise and identify their reductions  to coupled matching problems and 
primal-dual cycle summation problems. 
A detailed discussion of the depolarizing channel, related prior work, and the resulting open
problems is deferred to Section~\ref{depolS}.

\subsection{Outline}

Section~\ref{Preliminaries} reviews CSS codes, describes the toric, planar surface, and rotated
surface codes, and reviews basic matching preliminaries.
Section~\ref{RedSec} develops the local reduction of SMW decoding to MWPM via decorated graphs,
including the degree-3-defect resolution theorems for the toric, planar, and rotated settings.
Section~\ref{planarSep} presents planar-separator–based algorithms for solving the resulting MWPM
instances.
Section~\ref{FKT-SMLC} presents the SMLC decoder.  
Section~\ref{depolS} discusses extensions to the depolarizing channel.
Section~\ref{extAndOpenQuestions} collects further extensions and open questions.
The proofs of the degree-3-defect resolution theorems are given in
Appendices~\ref{AppendixA} and~\ref{AppendixB}. 
Appendix~\ref{app:pfaffians} reviews the necessary preliminaries on Pfaffians and Pfaffian orientations.

\newpage 
\section{Preliminaries}\label{Preliminaries}
 
We review CSS codes at a general algebraic level in Section \ref{CSS} to 
formalize syndromes, errors, and cosets. Then 
 we describe the 
toric, planar, and rotated surface codes as CSS
codes using a homological language in 
 Sections \ref{toricS}, \ref{planarS}, and \ref{rotatedS}.    
This provides the unified framework needed for both SMW and SMLC decoding. Finally, we review matching preliminaries in Section~\ref{matchings}.

\subsection{CSS codes}\label{CSS}

We use the standard notation for the single-qubit 
Pauli operators \(I,X,Y,Z\) and the 
\(n\)-qubit \emph{Pauli group} \(\mathcal{P}_n\),
consisting of all \(n\)-fold tensor products of the 
single-qubit Pauli matrices \(I, X, Y, Z\),
scaled by phases \(\pm 1\) and \(\pm i\). 
We assume familiarity with the stabilizer formalism and basic notions
of quantum error correction; see Nielsen and Chuang~\cite{nielsen-chuang}
for a standard reference.

A \emph{CSS (Calderbank–Shor–Steane) code} 
\cite{CalderbankShor1996,Steane1996} is a stabilizer code \cite{Gottesman1997} whose 
stabilizer admits a generating set that splits into two sets of 
\(X\)-type and \(Z\)-type operators.
 Equivalently, the stabilizer \(S\) is determined by two classical 
$\F_2$-linear codes \(C_X, C_Z \subset \mathbb{F}_2^n\) satisfying 
\(C_X^\perp \subset C_Z\).  
In particular, \(S\) is generated by \(X\)-type operators associated with a basis 
of \(C_X^\perp\) and \(Z\)-type operators associated with a basis of \(C_Z^\perp\). These bases are the rows of parity-check matrices 
\(H_X\) and \(H_Z\)  of \(C_X\) and \(C_Z\), 
respectively.

The parameters of a CSS code are  \( [[n,k,d]] \), where $n$ is the number of physical  qubits and  
$k = \dim C_X / C_Z^{\perp}$
is the number of logical qubits.  
 The code minimum distance $d = \min\{ d_X , d_Z \}$, 
where $d_X$ and $d_Z$ denote the minimum Hamming weights of any nontrivial logical $X$-type and $Z$-type operator, respectively; that is, they  are the minimum Hamming weights of a codeword in 
$C_X \setminus  C_Z^\perp$
 and $C_Z \setminus  C_X^\perp$, respectively.

An \(X\)-type Pauli error is specified entirely by its support, which we 
represent by a binary vector \(e_X \in \mathbb{F}_2^n\), and similarly a 
\(Z\)-type error is specified by a vector \(e_Z \in \mathbb{F}_2^n\).
In the decoding problem, we are given an \(X\)-syndrome 
\(s_X \in \mathbb{F}_2^{\dim(C_X^\perp)}\) and a \(Z\)-syndrome 
\(s_Z \in \mathbb{F}_2^{\dim(C_Z^\perp)}\). 
These syndromes record the parity of the overlap between the error support and the supports of the stabilizer generators.
Thus the error vectors must satisfy the CSS syndrome equations $H_Z e_X = s_Z$ and 
$H_X e_Z = s_X$.
The goal of decoding is to find error vectors \((e_X,e_Z)\) consistent with 
the observed syndromes. 

Throughout most of this paper, we work within the standard independent \(X/Z\) noise
model, where \(X\)- and \(Z\)-type errors are generated independently according to
binary symmetric channels with parameter \(p\). 
Under this assumption, CSS decoding separates into
two independent problems.   
Within this model, we consider two standard decoding objectives:

\begin{itemize}
 \item {\em Separate Minimum Weight (SMW) decoding.} 
 Find \(e_Z\) satisfying \(H_X e_Z = s_X\) that has maximum probability,  
      i.e., minimum Hamming weight, and independently find 
      \(e_X\) satisfying \(H_Z e_X = s_Z\) that has maximum probability,  i.e., 
      minimum Hamming weight.

\item  {\em 
Separate Most Likely Coset (SMLC) decoding.}
      Find \(e_Z\) such that \(H_X e_Z = s_X\) and the probability of the 
      coset \(e_Z + C_Z^\perp\) is maximized, and independently find 
      \(e_X\) such that \(H_Z e_X = s_Z\) and the probability of the 
      coset \(e_X + C_X^\perp\) is maximized.
	 
\end{itemize}
Coset-level decoding accounts for degeneracy, since errors differing by
stabilizers yield identical syndromes.

A more realistic physical noise model is the depolarizing channel, which does 
not assume independence between \(X\)-type and \(Z\)-type errors.  
Decoding under the depolarizing model is significantly more difficult, and no optimal 
polynomial-time  decoding algorithms are known for surface 
codes (see Section \ref{depolS}).

Surface codes are CSS codes whose stabilizer structure is induced by
two-dimensional cell complexes. In this setting, $X$-type checks correspond
to vertices, $Z$-type checks correspond to plaquettes, errors are represented as
$1$-chains, and syndromes are their boundaries.
In what follows, 
we describe these constructions explicitly for the toric code, its
planar realization, and the rotated surface code.

\subsection{The toric code}\label{toricS}

\begin{figure}[H]
  \centering
    \includegraphics[width=0.5\textwidth]{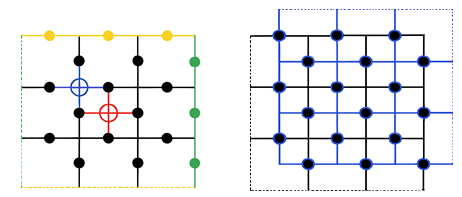}
\caption{%
{\bf Left:} The toric code lattice $\mathcal{X}$ for $L=3$.  
  The orange edges on the lower side and their incident vertices are identified 
  with those on the upper side.  Likewise, the green edges on the left side are 
  identified with those on the right side.  
   One  $X$-type check is  shown in blue and one $Z$-type check in red.  
  Two noncontractible cycles $a$ and $b$ of minimum length that are not 
  homotopic to each other are shown in orange and green.  
  {\bf Right:} The dual lattice $\mathcal{X}^\ast$ is shown in blue.}
  \label{toric}
\end{figure}

The {\em toric code} \cite{Kitaev1997} 
 is a $[[2L^2, 2, L]]$
CSS code built from an $L \times L$ cellulation $\mathcal{X}$  
of the torus, which we also refer to as a lattice. 
Associate one qubit with each of the $n = 2L^2$ edges, an $X$-type
check with each vertex, and a $Z$-type check with 
each plaquette, as shown in Figure~\ref{toric}. 

Thus $C_X^\perp$ and $C_Z^\perp$ are spanned 
by the vertex checks and the plaquette checks, 
respectively.
Consider the {\em dual lattice} $\mathcal{X}^\ast$ of $\mathcal{X}$, whose
vertices correspond to plaquettes of the primal and vice versa.
There is a one-to-one correspondence between the edges of $\mathcal{X}$ and
those of $\mathcal{X}^\ast$: they are identified whenever they intersect.

Let $F_i(\mathcal{X})$ be the set of $i$-faces of $\mathcal{X}$, and let
$C_i(\mathcal{X}) = \mathbb{F}_2[F_i(\mathcal{X})]$ denote the space of 
$i$-chains for $i = 0,1,2$.  
Thus $F_0$, $F_1$, and $F_2$ correspond to 
vertices, edges, and plaquettes, respectively.
Likewise, define $F_i(\mathcal{X}^\ast)$ and 
$C_i(\mathcal{X}^\ast)$ for the dual lattice.
We have the {\em chain complex} over $\mathbb{F}_2$:
\[
0 \xrightarrow{0} 
C_{2}(\mathcal{X}) \xrightarrow{\partial_{2}} 
C_{1}(\mathcal{X}) \xrightarrow{\partial_{1}} 
C_0(\mathcal{X}) \xrightarrow{0} 0,
\]
where the boundary maps $\partial_2$ and $\partial_1$ are defined by:
the boundary of a plaquette is the sum over $\F_2$ of its four edges, and 
the boundary of an edge is the sum over 
$\F_2$ of its two endpoints. See Hatcher~\cite{Hatcher} for a general reference on cellular homology.

In these terms, 
$C_X = \text{Ker}~\partial_1$ consists of 
     cycles in the primal lattice,  and 
$C_Z^\perp = \text{Im}~\partial_2$
consists of 
 boundaries in  the primal lattice.
The fact that $C_Z^\perp \subset C_X$ is captured by the chain complex property 
$\partial_1 \circ \partial_2 = 0$, i.e., every boundary is a cycle.

Similarly, the dual lattice supports the chain complex
\[
0 \xrightarrow{0} C_{2}(\mathcal{X}^\ast) \xrightarrow{\partial_{2}^\ast} C_{1}(\mathcal{X}^\ast)
 \xrightarrow{\partial_{1}^\ast} C_0(\mathcal{X}^\ast) \xrightarrow{0} 0, 
\]
where 
$C_Z = \text{Ker}~\partial_1^\ast$ consists of 
   cycles in the dual lattice, and 
$C_X^\perp = \text{Im}~\partial_2^\ast$  consists of   boundaries in the dual lattice.

The first homology group $H_1(\mathcal{X}) = C_X / C_Z^\perp$
captures the logical Pauli operators.
It is a $2$-dimensional vector space generated by 
${a} + C_Z^\perp$ and ${b} + C_Z^\perp$, where 
${a}$ and ${b}$ are two 
noncontractible primal cycles of minimum length that are not homotopic to
each other, as illustrated in Figure~\ref{toric}.
The same applies to the dual lattice.
The code minimum distance is the length $L$ of the shortest representatives, and it is   
 achieved by ${a}$, 
${b}$, and their dual analogues.

The $X$-syndrome is a primal $0$-chain 
$s_X \in 
C_0(\mathcal{X})$, and 
the $Z$-syndrome is a dual $0$-chain 
$s_Z \in C_0(\mathcal{X}^\ast)$.
We call the supports of $s_X$ and  $s_Z$ 
 $X$- and $Z$-{\em syndrome defects}, respectively. 
Since the sum  of
 all $X$-type checks is zero, the syndrome representation 
 is redundant.
If one wants to avoid redundancy, one vertex should be removed 
from $F_0(\mathcal{X})$, i.e., 
 the rows of $H_X$ should be indexed by all but one primal vertex.
The same applies to $F_0(\mathcal{X}^\ast)$. However,  
it is convenient not to remove any vertices and instead assume
that $s_X$ and $s_Z$ have even Hamming weight.  
Under this assumption, each syndrome is the boundary of some error.
Thus the errors  we wish to find are a {\em primal 
 $1$-chain}  
$e_Z$ and a {\em dual 
 $1$-chain}   $e_X$  whose boundaries are the syndromes: 
$\partial_1 e_Z = s_X$, 
and 
$\partial_1^\ast e_X = s_Z$.

\subsection{The planar surface code}\label{planarS}

The toric code is defined on a periodic lattice, which is difficult to realize in 
physical architectures.  This motivates the use of planar codes \cite{BravyiKitaev1998, Dennis2002}, where boundaries 
are introduced to enable implementation on flat  two-dimensional layouts.

\begin{figure}
  \centering
    \includegraphics[width=0.7\textwidth]{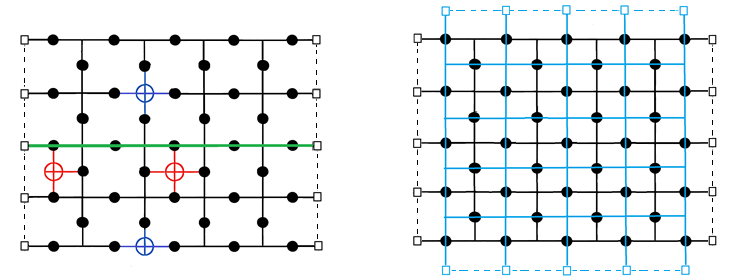}
  \caption{
  {\bf Left:} The cell complex $\mathcal{X}$ of the $L\times (L+1)$ rectangular 
  lattice for $L=5$. The edges of its $1$-dimensional subcomplex $\mathcal{A}$ 
  are shown with dashed lines, and the vertices of $\mathcal{A}$ are indicated 
  by boxes. Two $X$-type checks are shown in blue and two $Z$-type checks in 
  red. A minimal-length relative cycle $a$ connecting a vertex on the left side 
  of $\mathcal{A}$ to a vertex on the right side is shown in green.  
  {\bf Right:} The dual lattice is shown in blue.}
  \label{planar}
\end{figure}

Consider the cell complex $\mathcal{X}$ of the $L\times (L+1)$ rectangular 
lattice, and its $1$-dimensional subcomplex $\mathcal{A}$ consisting of the left 
and right side edges together with their incident vertices, as shown in 
Figure~\ref{planar}
(We will  refer to $\mathcal{X}$ interchangeably as a lattice or a cell complex, and similarly to $\mathcal{A}$ as a sublattice or a subcomplex, depending on the context.) 
The \emph{planar surface code} is the  
$[[2L^2 - 2L + 1,\,1,\,L]]$ CSS code obtained by associating a qubit with each 
edge not contained in $\mathcal{A}$, an $X$-check with each vertex not 
in $\mathcal{A}$, and a $Z$-check with each plaquette.

Relative homology provides a clean description of the code \cite{Bombin2007B,GTA2022}.
For $i=0,1,2$, define the space of \emph{relative $i$-chains} by modding out by  
the chains supported on $\mathcal{A}$:
$C_i(\mathcal{X},\mathcal{A})
:= C_i(\mathcal{X}) / C_i(\mathcal{A})
= \mathbb{F}_2[F_i(\mathcal{X}) \setminus F_i(\mathcal{A})]$. 
This yields the chain complex:
\[
0 \xrightarrow{0}
C_2(\mathcal{X},\mathcal{A})
\xrightarrow{\partial_2}
C_1(\mathcal{X},\mathcal{A})
\xrightarrow{\partial_1}
C_0(\mathcal{X},\mathcal{A})
\xrightarrow{0} 0 .
\]
Here 
$C_X = \operatorname{Ker} \partial_1$ consists of relative cycles,
$C_Z^\perp = \operatorname{Im}\partial_2$ consists of relative boundaries,
and $C_Z^\perp \subset C_X$ follows from $\partial_1 \circ \partial_2 = 0$.

Consider the dual cell complex $\mathcal{X}^\ast$ and its $1$-dimensional 
subcomplex $\mathcal{A}^\ast$,  as shown in Figure~\ref{planar}. 	
They are isomorphic to the primal complex and its subcomplex, up to rotation. Under this duality, there is a one-to-one correspondence between
the plaquettes of \(\mathcal{X}\) and the vertices of \(\mathcal{X}^\ast \setminus  \mathcal{A}^\ast\);
the edges of \(\mathcal{X} \setminus \mathcal{A}\) and the edges of \(\mathcal{X}^\ast \setminus \mathcal{A}^\ast\);
and the vertices of \(\mathcal{X}\setminus\mathcal{A}\) and the plaquettes of \(\mathcal{X}^\ast\).
   Associated with the dual is the chain complex:
\[
0 \xrightarrow{0}
C_2(\mathcal{X}^\ast,\mathcal{A}^\ast)
\xrightarrow{\partial_2^\ast}
C_1(\mathcal{X}^\ast,\mathcal{A}^\ast)
\xrightarrow{\partial_1^\ast}
C_0(\mathcal{X}^\ast,\mathcal{A}^\ast)
\xrightarrow{0} 0 ,
\]
where 
$C_Z = \text{Ker}~\partial_1^\ast$ consists of dual relative cycles, and  
$C_X^\perp = \text{Im}~\partial_2^\ast$ consists of dual relative 
boundaries.

The first relative homology group $H_1(\mathcal{X},\mathcal{A}) = C_X / C_Z^\perp$
captures the logical Pauli operators.
It is a $1$-dimensional vector space generated by $a + C_Z^\perp$, where $a$ is 
the shortest relative cycle connecting a vertex on the left side of $\mathcal{A}$
to a vertex on the right side, as illustrated in Figure~\ref{planar}.  
 The same  holds for the dual lattice.  
The minimum distance of the code is the length $L$ of the shortest such 
representatives, achieved by $a$ and its dual analogue.

The $X$-syndrome is a primal relative $0$-chain 
$s_X \in C_0(\mathcal{X},\mathcal{A})$, and
the $Z$-syndrome is a dual relative $0$-chain 
$s_Z \in C_0(\mathcal{X}^\ast,\mathcal{A}^\ast)$.
Unlike the toric code, the $X$-type checks are independent and so are $Z$-type checks:
$H_X \simeq \partial_1$ and $H_Z \simeq \partial_1^\ast$. Thus the errors we seek,
$e_Z \in C_1(\mathcal{X},\mathcal{A})$ and 
$e_X \in C_1(\mathcal{X}^\ast,\mathcal{A}^\ast)$,
are the \emph{relative $1$-chains} whose relative boundaries are the observed syndromes:
$\partial_1 e_Z = s_X$ and 
$\partial_1^\ast e_X = s_Z$.

\subsection{The rotated surface code}\label{rotatedS}

By working with a rotated lattice~\cite{Bombin2007,rotated2}, the parameters of the planar 
surface code can be improved to $[[L^2,\,1,\,L]]$.

\begin{figure}[H]
  \centering
    \includegraphics[width=0.8\textwidth]{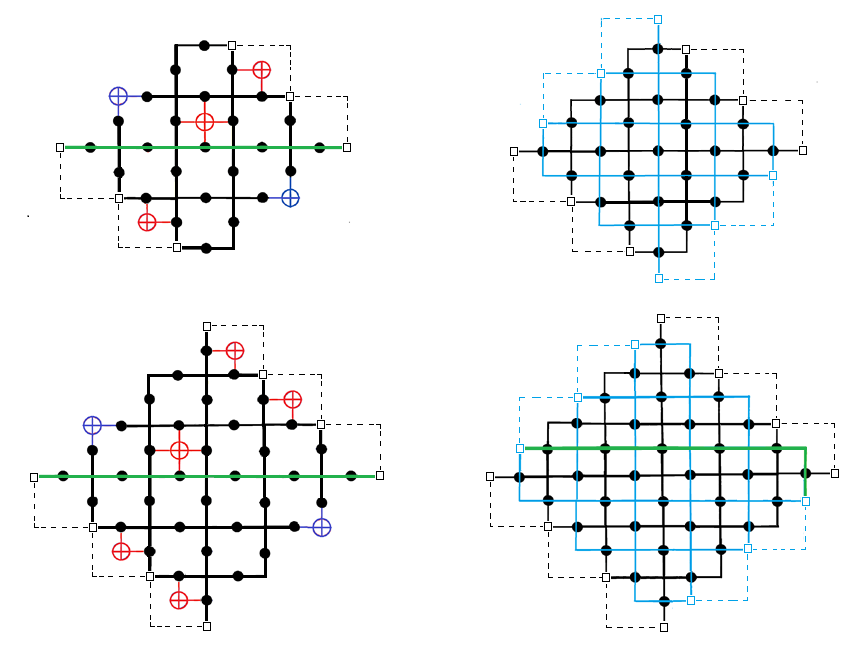}
\caption{   
The {\bf top row} shows a rotated surface code for odd $L$:  $L = 5$.  
The {\bf bottom row} shows a rotated surface code for even $L$: $L = 6$.
The cell complex $\mathcal{X}$ and its subcomplex $\mathcal{A}$ are shown on 
the left. The edges of $\mathcal{A}$ are drawn with dashed lines.  
The vertices of $\mathcal{A}$ are the endpoints of these edges.  
Among them, the vertices indicated by boxes are those incident to edges 
outside $\mathcal{A}$; these are the ones relevant for decoding.
A minimal-length relative cycle $a$ that is not a relative boundary is shown 
in green.
The dual lattice is shown in blue on the right. 
  }
  \label{rotated}
\end{figure}

Consider the cell complex $\mathcal{X}$ of the width-$L$ diamond-shaped lattice 
and the associated $1$-dimensional subcomplex $\mathcal{A}$, as shown in 
Figure~\ref{rotated}.  
The rotated surface  code is defined with respect to $(\mathcal{X},\mathcal{A})$ 
exactly as in the previous section.

A key technical difference is that some vertices in $\mathcal{A}$ are incident 
only to edges in $\mathcal{A}$.  
Such vertices are irrelevant for decoding.  
The vertices of $\mathcal{A}$ that matter are precisely those incident to at 
least one edge of $\mathcal{X}\setminus \mathcal{A}$; these vertices are indicated by boxes in the figure.

When $L$ is odd, the dual complex and its subcomplex 
are isomorphic to the primal complex and its subcomplex,  up to rotation, as 
illustrated in Figure~\ref{rotated}.  
When $L$ is even, the dual has a slightly different structure, but it  
still achieves the same minimum distance~$L$, as shown in the figure.

\subsection{Matchings}\label{matchings}

A \emph{matching} in an undirected graph \(G \) is a subset of edges in which no two edges share a vertex.  
A \emph{perfect matching} is a matching that covers every vertex of \(G\); that is, each vertex of \(G\) is incident to exactly one edge of the matching (hence the number of vertices of $G$ must be even).  
If the graph is weighted, the weight of a matching \(M\) is the sum of the weights of its edges.
In the \emph{Minimum Weight Perfect Matching} (MWPM) problem, we are given an edge-weighted undirected graph that is guaranteed to admit a perfect matching, and the goal is to find a perfect matching \(M\) of minimum weight.
In a seminal 1965 paper~\cite{Edmonds1965}, Edmonds introduced the blossom algorithm, which solves the MWPM problem in polynomial time.
See \cite{LovaszPlummer} for a general reference on Matchings.

\section{Local reduction of SMW to MWPM}\label{RedSec}

Consider the SMW decoding problem for the toric code.  
Given a primal $0$-chain $s_X$ and a dual $0$-chain $s_Z$, representing the
$X$- and $Z$-syndromes respectively, we seek minimum weight error chains.  
More precisely, we wish to find a
minimum weight 
 primal $1$-chain $e_Z$ whose boundary is the
$X$-syndrome,
$\partial_1 e_Z = s_X$,
and independently a minimum weight dual $1$-chain $e_X$ whose dual boundary is the $Z$-syndrome,
$\partial_1^\ast e_X = s_Z$. 

Consider the problem of decoding $e_Z$. In prior work, beginning with the seminal paper~\cite{Dennis2002},  this is  reduced to a
MWPM problem as follows.
Construct the complete \emph{$Z$-decoding graph} 
 whose vertices are the
\emph{$X$-syndrome defects}, and assign to each
pair $(u,v)$ of vertices a weight equal to the length of the shortest path from
$u$ to $v$ in the primal lattice $\mathcal{X}$.  
Then the decoding problem reduces to finding a  MWPM in the  $Z$-decoding graph, which can be solved using Edmonds' blossom algorithm
\cite{Edmonds1965}.  
Decoding $e_X$ proceeds analogously on the \emph{$X$-decoding graph} built
from $Z$-syndrome defects in the dual lattice.
Planar and rotated surface codes can be decoded similarly by introducing
appropriate virtual vertices into the decoding graphs. 

The worst-case complexity of solving the resulting MWPM problem using the blossom algorithm is
$O(n^{3} \log n)$~\cite{HiggottGidney2023,Higgott2022}.   

We propose a different reduction of the SMW decoding problem to the MWPM problem, yielding a substantial improvement in decoding time. Instead of reducing SMW decoding to the complete $X$- and $Z$-decoding graphs, we reduce it to instances of MWPM on sparser graphs, called the \emph{decorated graphs}, one for decoding $X$ errors and one for decoding $Z$ errors.
The decorated graphs are constructed from the primal  lattice and its dual, respectively, by replacing each vertex of the lattice  with a small gadget.
The gadgets encode the parity constraints imposed by the syndrome and are based on the work of Fisher~\cite{Fisher66}.
In the planar and rotated surface codes, they are also used to encode boundary conditions. 
The reduction preserves the local structure of the lattice. In the toric code case, the resulting graph has genus~$1$, while in the planar and rotated surface code cases it is planar. This structural property enables the use of planar-separator-based methods due to Lipton and Tarjan~\cite{LiptonTarjan80} to efficiently solve the resulting MWPM instances. The improvement in decoding time arises from this structural property together with the reduced size of the decoding graph.
We further optimize the size of the decorated graph by allowing limited local nonplanarity, while retaining the applicability of planar-separator-based methods. In this section, we describe the reduction, while in Section~\ref{planarSep} we describe  the corresponding efficient MWPM solvers.

We start with the reduction for the toric code in Section~\ref{ToricReduction}, and then adapt it to planar surface codes and rotated surface codes in Sections~\ref{redPlanar} and~\ref{redRot}, respectively.

\subsection{The toric code}\label{ToricReduction}
The SMW decoding problem for the toric code can be naturally expressed using $D$-joins.  
If $G$ is an undirected graph and $D$ is a subset of vertices of  
$G$ of even size, a {\em $D$-join} is a set $J$ 
of edges of $G$ 
such that the $J$-degree of each vertex in $D$ 
is odd and the $J$-degree of each vertex of $G$ outside $D$ is even,  
where the $J$-degree of a vertex $v$ of $G$ is the number of edges in $J$ incident to $v$.

In terms of \(D\)-joins, the problem of decoding \(e_Z\) can be phrased as follows.  
Let \(G\) be the toric-code lattice viewed 
as a genus-$1$ graph  and \(D\) the set of \(X\)-syndrome \emph{defects}, i.e., 
the support of \(s_X\).  Thus the size of $D$ is even since the sum of all $X$-checks is even.
A primal \(1\)-chain \(e_Z\) whose boundary is \(s_X\) is a \(D\)-join.  
Thus, we need to solve the {\em Minimum Size \(D\)-Join (MSJ) problem}.

The reduction from MSJ to MWPM  goes back to Edmonds’s matching theory 
(see Exercise~8.0.4 in the monograph of Lov\'asz and Plummer~\cite{LovaszPlummer}).  
The idea is  simple: each vertex in \(D\) is replaced by a gadget that forces an odd number of 
incident matching edges to be used, while each vertex outside \(D\) is replaced by a gadget that 
forces an even number of incident matching edges to be used.  
These local parity-enforcing gadgets originate in the classical work of Fisher in 1966~\cite{Fisher66} in his study of the partition function of the planar Ising model.
Fisher’s gadgets are precursors of what later became known as \emph{matchgates} \cite{Valiant2002}.

\begin{figure}[H]
  \centering
    \includegraphics[width=1\textwidth]{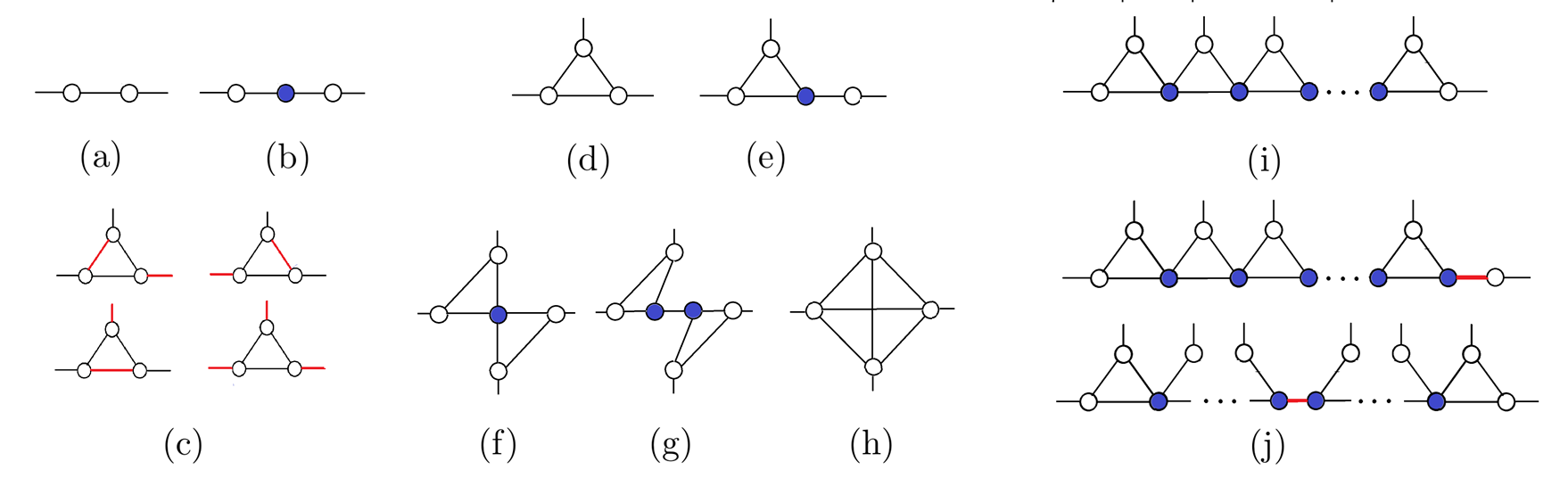}
\caption{
(a) \textbf{Even degree-2 Fisher gadget}.
(b) \textbf{Odd degree-2 Fisher gadget}: It has one 
internal vertex  shown in blue.
(d) \textbf{Odd degree-3 Fisher gadget}: Any perfect matching requires an odd number of vertices  to be 
matched externally, as shown in (c) where 
the perfect matching edges are shown in red.  
(e) \textbf{Even degree-3 Fisher gadget}: 
Introducing the additional edge flips the parity.  The gadget has one 
internal vertex  shown in blue.  
(f) \textbf{Odd degree-4 Fisher gadget}: It is constructed by gluing together two odd degree-3 gadgets along the internal vertex shown in blue.  
(g) \textbf{Even degree-4 Fisher gadget}: Obtained by adding an edge that flips the parity.  
(h) \textbf{Even nonplanar degree-4 gadget}: Although not planar, it behaves like the gadget in (g) and has smaller size.  
Its complete internal structure allows excluding any even number of vertices from the internal matching, thereby forcing an even number of  vertices to be externally matched.  
(i) \textbf{Odd degree-$k$ Fisher gadget}, for \(k \ge 3\): It is obtained by gluing \(k-2\) odd degree-3 gadgets. It  follows by induction that 
this gadget has odd parity.     
The gadget has \(k-3\) internal vertices  shown in blue and a total of \(3(k-2)\) internal edges.  
(j) \textbf{Even degree-$k$ Fisher gadget}, for \(k \ge 3\): It is obtained by adding a parity-flipping edge to the odd degree-$k$ gadget.  
It has \(k-2\) internal vertices and \(3k-5\) internal edges. The parity-flipping edge is shown in red. 
It may be added on the boundary or in the interior of the gadget, as illustrated by two realizations of the gadget.
}
\label{fisher}
\end{figure}

In what follows we define the gadgets needed in this paper. 
A \emph{gadget} is a graph whose vertices are partitioned into two types: 
\emph{external} and \emph{internal}.  
The \emph{degree} of a gadget is the number of its external vertices.  
We label a gadget as \emph{odd} (resp.\ \emph{even}) if, for any odd (resp.\ even) subset of its external vertices, 
the graph obtained by excising those vertices has a unique perfect matching.
Thus an even (resp.\ odd) gadget forces any odd (resp.\ even) number of external vertices to be matched externally. Figure~\ref{fisher} shows the even and odd degree-\(k\) Fisher gadgets for \(k \ge 2\).
All gadgets shown are planar except for one degree-$4$ gadget in Part~(h), which we consider in this section due to its low complexity.  
Although only degree-$4$ gadgets are needed in this section, we include the general degree-$k$ construction for future reference in Section~\ref{redPlanar}.

Thus, the minimum size \(D\)-join problem reduces to a MWPM instance on the 
{\em decorated graph} \(G_D\), obtained from \(G\) and \(D\) as follows.
Replace each vertex in \(D\) by an odd degree-4 Fisher gadget, and each 
vertex not in \( D\) by an even degree-4 Fisher gadget.  
Connect the external vertices of the gadgets according to the adjacency 
structure of \(G\).  
Assign weight \(0\) to all gadget-internal edges and weight \(1\) to all edges 
inherited from \(G\).
This construction yields a weight-preserving one-to-one correspondence 
between \(D\)-joins of \(G\) and perfect matchings of \(G_D\).

The size of the decorated graph is linear in the size of the lattice, 
which already provides an advantage over constructing the complete 
decoding graph on the syndrome defects.
Moreover, \(G_D\) has genus~\(1\), like the underlying torus, 
which enables the use of planar separator methods to solve the MWPM 
instance more efficiently, as explained  
in  Section~\ref{planarSep}.

To further reduce the size of the decorated graph, we may use the 
even degree-4 \emph{nonplanar} gadget for vertices outside \(D\).  
Although this increases the genus, planar separator 
methods can be adapted to handle such localized nonplanarity.  

The size of the decorated graph can be reduced even further by 
omitting the odd  gadgets at defect vertices entirely, and using instead one vertex per defect as  in 
the original graph. 
If \(J\) is a \(D\)-join of \(G\), then each 
vertex in \(D\) has \(J\)-degree \(1\) or \(3\).
We show in Theorem \ref{DJoin} that there is a minimum size \(D\)-join in which all vertices in \(D\) have \(J\)-degree \(1\). 
We verify this under the assumption that the distance 
\(L\) of the  toric code is even.  
We believe the evenness assumption is unnecessary, but our current proof 
does not cover the odd-\(L\) case.

\begin{theorem}[Dissolving degree-$3$ defects in the toric code] \label{DJoin}
Let \(G = (V,E)\) be the graph of the lattice   
(or dual lattice) $\mathcal{X}$  of the 
\([[2L^2,2,L]]\) toric code.  
Assume that \(L\) is even and \(L \ge 4\).
Let \(D \subset V\) be a set of even cardinality.
Then \(G\) has a minimum-size \(D\)-join \(J\) such that 
every vertex in \(D\) has \(J\)-degree \(1\).

More specifically, if \(J\) is a minimum-size \(D\)-join of \(G\), 
then there exists a \(2\)-chain \(c\) in $\mathcal{X}$   such that 
\(J' = J \oplus \partial c\) satisfies \(|J'| = |J|\) and 
every vertex in \(D\) has \(J'\)-degree \(1\).
\end{theorem}

Note that \(J\) and \(J'\) belong to the same homology class.
We establish the theorem by iteratively resolving  defects with $J$-
degree~$3$  by 
xoring \(J\) with boundaries of plaquettes, until all defect 
vertices have degree~1.  
The proof is long and technical in nature, and we defer it to  Appendix \ref{AppendixA}.

\begin{figure}[H]
  \centering
    \includegraphics[width=0.5\textwidth]{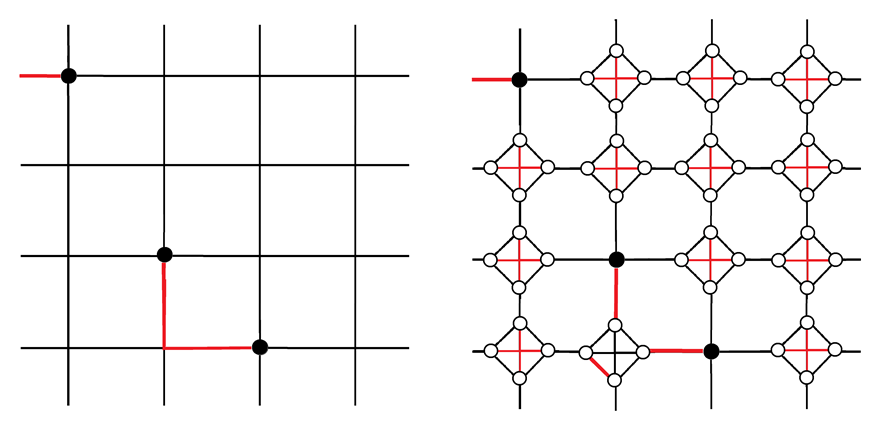}
\caption{  
{\bf Left:} A portion of the graph \(G\) of the toric code, where the \(X\)-syndrome defects are shown in filled black circles. The $D$-join edges are shown in red. 
{\bf Right:} the decorated graph $G_D$ using one vertex per defect in \(D\), and the even degree-4 nonplanar gadget for vertices outside \(D\).  The perfect matching edges are shown in red. 
}
\label{decoratedTorus}
\end{figure}

See Figure~\ref{decoratedTorus} for an illustration of the reduction.

The problem of decoding \(e_X\) can be reduced to a MWPM problem on a decorated graph derived from the dual lattice and the \(Z\)-syndrome $s_Z$ in exactly the same way.

\subsection{The  planar surface code}\label{redPlanar}

To adapt the reduction to planar and rotated surface codes, we define the notion of  relative $D$-joins.

Let $G = (V,E)$ be an undirected graph, and let $B$ and $D$ be disjoint subsets of the vertex set of $G$.  
A \emph{$D$-join of $G$ relative to $B$} is a set $J$ of edges of $G$ such that the $J$-degree of each vertex in $D$ is odd, and the $J$-degree of each vertex in $V\setminus (D\cup B)$ is even.

While the $J$-degrees of the vertices in $B$ are not individually constrained, the total $J$-parity on $B$ is determined by the cardinality of $D$.  
Namely, the parity of $|D|$ equals the \emph{total $J$-parity} of $B$, where the latter is defined as the sum modulo $2$ of the $J$-degrees of the vertices in $B$.  
This follows from the fact that the sum of the $J$-degrees over all vertices of $G$ must be even.

Consider the planar surface code with parameters 
$[[2L^2 - 2L + 1,\,1,\,L]]$, as defined in Section~\ref{planarS}, in terms of the 
$L\times (L+1)$ lattice $\mathcal{X}$ and its $1$-dimensional sublattice $\mathcal{A}$. 

We consider first the problem of decoding the error $e_Z$.
Given a relative $0$-chain $X$-syndrome $s_X \in C_0(\mathcal{X},\mathcal{A})$, 
the goal is to find a minimum weight relative $1$-chain 
$e_Z \in C_1(\mathcal{X},\mathcal{A})$ whose relative boundary is $s_X$, i.e., $\partial_1 e_Z = s_X$.

In terms of relative joins, the problem of decoding \(e_Z\) can be phrased as follows.
Let $G$ be the graph of $\mathcal{X}$ with the edges of $\mathcal{A}$ removed (i.e., excluding the dashed lines in Figure~\ref{planar}), and let $B$ be the set of vertices of $\mathcal{A}$ (the boxed vertices in Figure~\ref{planar}).  
Let $D$ be the set of relative $X$-syndrome defects, i.e., the support of $s_X$.

A relative $1$-chain whose relative boundary is $s_X$ corresponds to a $D$-join $J$ of $G$ relative to $B$.
Thus, decoding reduces to solving the minimum-size $D$-join problem relative to $B$, which we refer to as the \emph{MSRJ (Minimum Size Relative Join) problem}.

A key difference compared to the toric code is that here the cardinality of $D$ need not be even, since syndrome defects may be matched to vertices in $B$. As noted above, the parity of $|D|$ determines the total $J$-parity on $B$.

Following the toric code case, the MSRJ problem reduces to an MWPM instance on a \emph{decorated graph} \(G_D\), obtained from \(G\) and \(D\) as follows.

Each vertex in \(D\) is replaced by an odd Fisher gadget of degree $4$ or $3$ 
(depending on whether the vertex lies in the interior or on the boundary of the lattice), and each vertex not in \(D\cup B\) is replaced by an even Fisher gadget of degree $4$ or $3$.  

For the vertices in \(B\), we distinguish two cases depending on the parity of \(|D|\).  
{\em If \(|D|\) is even}, the vertices in \(B\) are 
 identified with the external vertices of a degree-$2L$ 
 even Fisher gadget.  
{\em If \(|D|\) is odd}, they are 
identified with the external vertices of a degree-$2L$ 
 odd Fisher gadget. 
  This choice enforces the constraint that the total  $J$-parity of the vertices in \(B\) must equal the parity of \(|D|\).

The external vertices of the gadgets are then connected according to the adjacency structure of \(G\).  
All gadget-internal edges are assigned weight \(0\), while edges inherited from \(G\) are assigned weight \(1\).

This construction yields a weight-preserving one-to-one correspondence between \(D\)-joins of \(G\) relative to \(B\) and perfect matchings of \(G_D\).

Analogously to the toric code setting, 
 we can reduce the size of the decorated graph by using the even degree-$4$ \emph{nonplanar}  gadget instead of its planar counterpart.  

Moreover, this reduction can be pushed further: as justified by Theorem~\ref{DJoinPlanar}, we may omit the odd  gadgets at defect vertices entirely and instead use a single vertex per defect, as in the original graph.

\begin{figure}[H]
  \centering
    \includegraphics[width=0.7\textwidth]{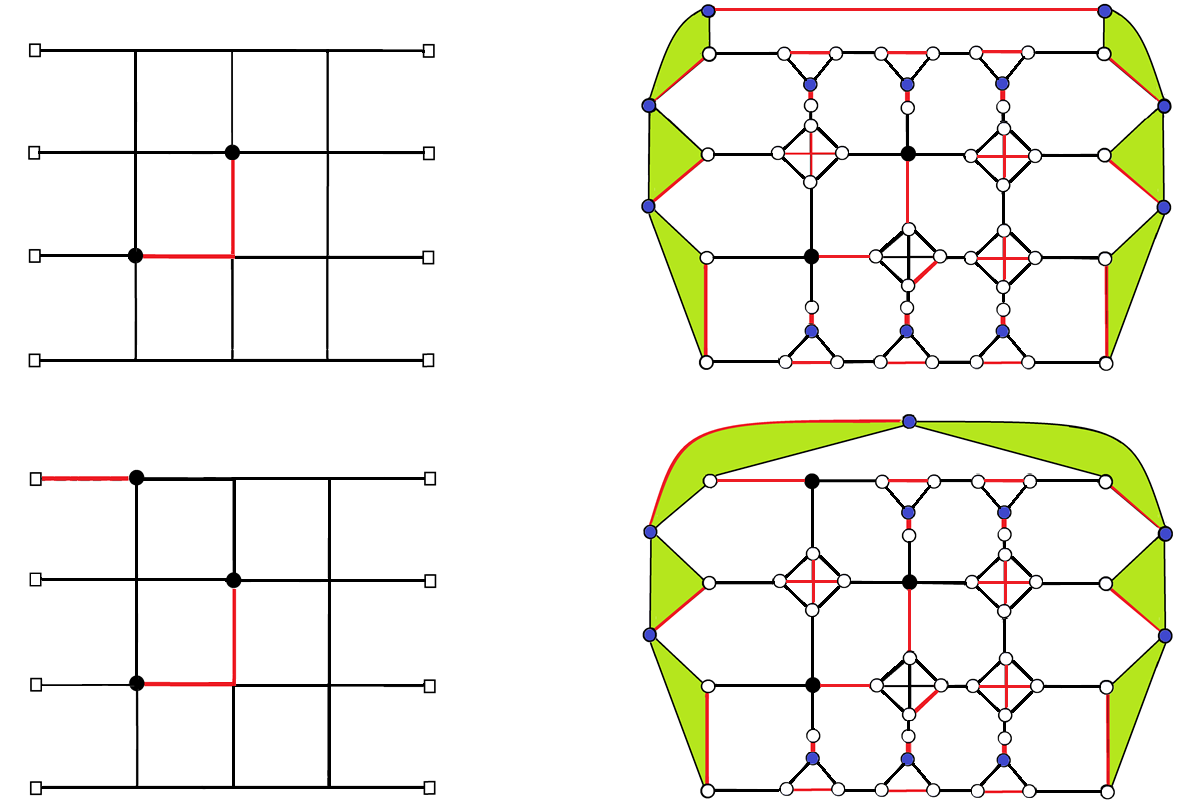}
  \caption{
  \textbf{Left:} The graph \(G\) of the planar surface code for \(L = 4\), where the vertices in \(B\) are shown as squares. 
  The vertices in \(D\) are shown as filled black circles, and the edges of a \(D\)-join relative to \(B\) are shown in red. 
  The \textbf{upper row} illustrates the case \(|D|\) even, while the \textbf{lower row} illustrates the case \(|D|\) odd. 
  \textbf{Right:} The corresponding decorated graph $G_D$, using one vertex per defect in \(D\), together with even degree-$4$ nonplanar  gadgets or planar degree-$3$ Fisher gadgets for vertices outside \(D\cup B\). 
  The gadget associated with \(B\) is highlighted using green triangles. 
  Vertices internal to the gadgets are shown in blue, and perfect-matching edges are shown in red.
  }	
\label{decoratedPlanar}
\end{figure}

See Figure~\ref{decoratedPlanar} for an illustration of the reduction.

\begin{theorem}[Dissolving degree-$3$ defects in the planar surface code]\label{DJoinPlanar}
Consider the planar surface code with parameters 
$[[2L^2 - 2L + 1,\,1,\,L]]$, as defined in Section~\ref{planarS}, in terms of the 
$L\times (L+1)$ lattice $\mathcal{X}$ and its $1$-dimensional sublattice $\mathcal{A}$. 
Let $B$ be the set of vertices of $\mathcal{A}$, and let $G = (V,E)$ denote the subgraph obtained from 
$\mathcal{X}$ by excluding the edges in $\mathcal{A}$. 
Assume that $L \geq 3$.

Let $D \subset V$ be a subset of vertices of $G$ disjoint from $B$. 
Then $G$ has a minimum-size $D$-join $J$ relative to $B$ such that every vertex in $D$ has $J$-degree $1$.

More specifically, if $J$ is a minimum-size $D$-join of $G$ relative to $B$, then there exists a $2$-chain $c$ in $\mathcal{X}$ such that, with
$J' = J \oplus \partial c$, 
where $\partial c$ denotes the relative boundary of $c$, we have $|J'| = |J|$ and every vertex in $D$ has $J'$-degree $1$.

The same statement holds if $G$ and $B$ are associated with the dual lattice $\mathcal{X}^\ast$ and its sublattice $\mathcal{A}^\ast$, as defined in Section~\ref{planarS}.
\end{theorem}

The proof of Theorem~\ref{DJoinPlanar} is given in Appendix~\ref{AppendixB}. 
Unlike the toric code case, the proof does not require the distance \(L\) to be even.

As in the toric code setting, the problem of decoding \(e_X\) is   reduced to a minimum weight perfect matching problem on a decorated graph derived from the dual lattice and the \(Z\)-syndrome $s_Z$ in exactly the same way.

\subsection{The rotated surface code}\label{redRot}

Consider the rotated surface code with parameters 
$[[L^2,\,1,\,L]]$, as defined in Section~\ref{rotatedS}, in terms of the width-$L$ diamond-shaped lattice $\mathcal{X}$ and its $1$-dimensional sublattice $\mathcal{A}$. Consider first  the problem of decoding $e_Z$:  given a relative $0$-chain $X$-syndrome $s_X$, 
 find a minimum weight relative $1$-chain
$e_Z $ whose relative boundary is $s_X$.

In terms of relative joins, the problem can be phrased as in the planar case, except that here the lattice $\mathcal{X}$ contains vertices that are incident only to edges in $\mathcal{A}$. 
Such vertices must be excluded from the graph $G$. Specifically, let $G$ be the graph of $\mathcal{X}$ obtained by removing the edges of $\mathcal{A}$ (i.e., the dashed lines in Figure~\ref{rotated}) and by deleting those vertices of $\mathcal{A}$ that become isolated. 
Let $B$ be the set of remaining vertices of $\mathcal{A}$ (the boxed vertices in Figure~\ref{rotated}).

Thus, we need to solve the minimum size \(D\)-join problem relative to \(B\). 
As in the planar surface code case, the MSRJ problem reduces to an MWPM instance on a \emph{decorated graph} \(G_D\), obtained from \(G\) and \(D\) as follows.

Each vertex in \(D\) is replaced by an odd Fisher gadget, and each vertex not in \(D\cup B\) is replaced by an even Fisher gadget.  
The degrees of the gadgets are now $2$ or $4$, depending on whether the corresponding vertices lie on the boundary or in the interior of the lattice.  
When the degree is $4$, a nonplanar gadget may be used to reduce the size of the decorated graph.

For the vertices in \(B\), all edges incident  to them are connected to a single Fisher gadget of degree $4L-O(1)$, whose parity is chosen to equal the parity of \(|D|\).  
As in the planar surface code case, the external vertices of the gadgets are connected according to the adjacency structure of \(G\).  
All gadget-internal edges are assigned weight \(0\), while edges inherited from \(G\) are assigned weight \(1\).

We can also reduce the size of the decorated graph by using the even degree-$4$ nonplanar  gadget instead of its planar counterpart.  
Moreover, as shown in Theorem~\ref{DJoinRotated}, we may omit the odd   gadgets at defect vertices entirely and instead use a single vertex per defect.

\begin{figure}[H]
  \centering
    \includegraphics[width=0.7\textwidth]{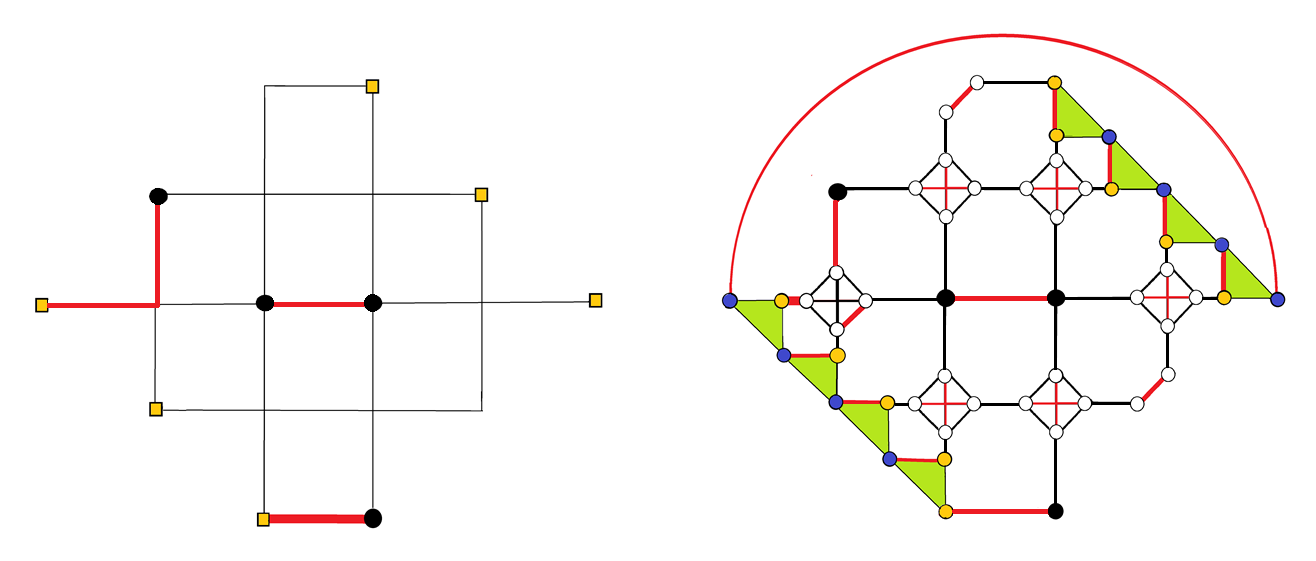}
\caption{
  \textbf{Left:} The graph \(G\) of the rotated surface code for \(L = 5\), where the vertices in \(B\) are shown as orange squares. 
  The vertices in \(D\) are shown as filled  black circles, and the edges of a \(D\)-join relative to \(B\) are shown in red. 
  This figure illustrates the case \(|D|\) even. 
  \textbf{Right:} The corresponding decorated graph $G_D$, using one vertex per defect in \(D\), together with even degree-$4$ nonplanar gadgets or planar degree-$2$ Fisher gadgets for vertices outside \(D\cup B\). 
  The gadget associated with \(B\) is highlighted using green triangles. 
  Vertices internal to the gadgets are shown in blue, and perfect-matching edges are shown in red.
  }
\label{decoratedRotated}
\end{figure}

See Figure~\ref{decoratedRotated} for an illustration of the reduction when the parity of $D$ is even.

\begin{theorem}[Dissolving degree-$3$ defects in the rotated surface code]\label{DJoinRotated}
Consider the rotated surface code with parameters 
$[[L^2,\,1,\,L]]$, as defined in Section~\ref{rotatedS}, in terms of the width-$L$ diamond-shaped lattice $\mathcal{X}$ and its $1$-dimensional sublattice $\mathcal{A}$.  
Assume that $L \geq 3$.

Let $G$ be the graph of $\mathcal{X}$ obtained by excluding the edges of $\mathcal{A}$ and deleting the vertices of $\mathcal{A}$ that become isolated, and let 
 $B$ be the set of  remaining vertices in 
 $\mathcal{A}$.

The claim of Theorem~\ref{DJoinPlanar} holds for this choice of 
 $\mathcal{X}, \mathcal{A}, G$ and $B$.  
The same statement also holds if $G$ and $B$ are associated with the dual lattice $\mathcal{X}^\ast$ and its sublattice $\mathcal{A}^\ast$, as defined in Section~\ref{rotatedS}.
\end{theorem}

The proof of Theorem~\ref{DJoinRotated} is given in Appendix~\ref{AppendixB}.

Decoding \(e_X\) proceeds analogously using the dual lattice and the \(Z\)-syndrome \(s_Z\).

\section{SMW decoding using planar separators}\label{planarSep}

The planar separator approach~\cite{LiptonTarjan80} is expressed in the framework of a variant of MWPM, namely the \emph{Maximum Weight Matching (MWM)} problem.  
In the MWM problem, given an edge-weighted undirected graph,
the goal is to find a (not necessarily perfect) matching \( M \) of maximum total weight. 
The MWPM and MWM problems are closely related, and it is folklore that each can be reduced to the other. 
Indeed, if \( G = (V,E) \) is a graph that is guaranteed to admit a perfect matching and is equipped with nonnegative edge weights
\( w : E \to [0,k] \), where \( k > 0 \), then an instance of MWPM on \( G \) can be reduced to an instance of MWM on the same graph
by defining modified edge weights \( w'(e) = k|V| - w(e) \), for all \( e \in E \).\footnote{
The modified weight of any non-perfect matching is at most
\((|V|/2 - 1)\,k|V|\), which is strictly less than the weight of any perfect matching, since the latter is at least
\((|V|/2)(k|V| - k)\).
}

Given a set \(D\) of \(X\)- or \(Z\)-syndrome defects, let \(G_D = (V,E)\) be the decorated graph constructed in the previous section for the toric code, the planar surface code, or the rotated surface code.  
In our setup, the original edge weights on \(G_D\) are either \(0\) or \(1\), so \(k=1\).  
Consequently, in the reduction from MWPM to MWM, the modified edge weights are either \(|V|\) or \(|V|-1\), depending on whether the edge is internal to a gadget or an external edge inherited from the underlying lattice. 
In what follows, let \(N\)  be the number of vertices  in \(G_D\).

Lipton and Tarjan gave an \(O(N^{3/2}\log N)\) algorithm~\cite{LiptonTarjan80} 
for the MWM problem on a planar graph with \(N\) vertices using planar separators. If, instead of using the nonplanar degree-4 gadgets in the construction of the decorated graph,  
we use the slightly larger planar degree-4 Fisher gadgets, then the decorated graph \(G_D\) 
becomes planar in the case of the planar surface code and the rotated surface code.  
By treating the Lipton–Tarjan algorithm as a black box,  
we immediately conclude that these codes can be decoded in 
\(O(N^{3/2}\log N)\) time. 
In Section~\ref{adaptation}, we tailor this algorithm to the special structure of these codes, which possess natural planar separators, and show how it can be adapted to accommodate both the local nonplanarity induced by nonplanar degree-$4$ gadgets and the global nonplanar structure of the toric code. 
We begin in Section \ref{GLTalg} with an overview of the planar separator approach to MWM.

\subsection{The Lipton--Tarjan planar separator algorithm}\label{GLTalg}

We start with definitions from matching theory following~\cite{LiptonTarjan80}. 
Let \(G\) be an edge-weighted undirected graph, and let \(M\) be a matching in \(G\).
An edge of \(G\) is called a \emph{matching edge} if it belongs to \(M\).
A vertex of \(G\) is called \emph{matched} if it is incident to a matching edge.
An \emph{alternating path}  in \(G\) with respect to \(M\) is a path or cycle whose edges alternate between matched and unmatched edges.
The \emph{net weight} of an alternating path is defined as the total weight of its unmatched edges minus the total weight of its matched edges. 
An \emph{augmenting path}  in \(G\) with respect to \(M\) is an alternating path of strictly positive net weight such that exchanging matched and unmatched edges along it yields another matching, i.e., it is either an alternating cycle, or an acyclic alternating path whose first and last vertices are not matched to  vertices outside the path.

The following lemma captures the main update step in Gabow's primal-dual
algorithm for MWM, which refines Edmonds' blossom framework.
\begin{lemma}[Finding an Augmenting Path~\cite{Gabow75,LiptonTarjan80}; \textnormal{see also}~\cite{GMG86}]\label{augPath}
Let \(G=(V,E)\) be an undirected edge-weighted graph, let \(v\in V\),
and let \(G\setminus v\) be the subgraph of \(G\) induced by the vertex set
\(V\setminus\{v\}\).
Suppose \(M\) is a maximum weight matching in \(G\setminus v\).
If \(G\) contains no augmenting path with respect to \(M\) having \(v\) as one
endpoint, then \(M\) is a maximum weight matching of \(G\).
Otherwise, let \(P\) be an augmenting path with respect to \(M\) having maximum
net weight.
Then the symmetric difference \(M \oplus P\) is a maximum weight matching of
\(G\).

Moreover, using appropriate data structures, both the existence test for such a path \(P\) and the construction of  one with maximum 
net weight  can be carried out in
\(O(|E|\log |V|)\) time.

\end{lemma}

We now turn to the {\em planar separator  theorem for planar graphs}.  
Lipton and Tarjan~\cite{LiptonTarjan79} showed that if \(G\) is a planar graph, then its vertex set can be partitioned into three sets \(A,B,C\) such that \(G\) has no edges connecting vertices in \(A\) to vertices in \(C\), \(|A|,|B| \le 2N/3\), and \(|C| \le 2^{3/2}\sqrt{N}\), where \(N\) is the number of vertices of \(G\).
Using this theorem, Lipton and Tarjan~\cite{LiptonTarjan80} proposed the following
divide-and-conquer algorithm for the MWM problem.
To solve the MWM problem on a planar graph \(G\),
apply the planar separator theorem to the vertices of \(G\),
and proceed recursively on the two subgraphs of $G$ induced on \(A\) and \(B\).
The subproblem solutions are then combined by applying
Lemma~\ref{augPath} at each vertex in the separator set \(C\).
The total running time is given by the recurrence
$$T(N) = 2T(2N/3) + O\!\left(|C|\,|E| \log N\right),$$
whose solution is \(T(N) = O(N^{3/2}\log N)\),
since \(|C| = O(\sqrt{N})\) and \(|E| = O(N)\) for planar graphs.

\subsection{Adaptation to an SMW decoder}\label{adaptation}

Rather than working with the atomic vertices of the decorated graph \(G_D\),
we operate at the level of the gadgets, which we refer to as \emph{gadget vertices}.
This allows us to handle the nonplanar structure that is confined within the even degree-$4$ gadgets. 
Thus, a gadget vertex is either a single vertex associated with a syndrome defect, or
an even gadget associated with a check vertex  that is not flagged as a defect in the measured syndrome 
(when \(L\) is odd in the toric code case, we use  an odd planar degree-4 Fisher gadget for each  defect, since the proof of Theorem~\ref{DJoin} applies only for even \(L\)). 
The large planar Fisher gadgets associated with the boundary vertices in the planar or rotated surface codes will be handled separately.

We tailor the Lipton--Tarjan planar separator algorithm to the toric code setting in Section~\ref{tccPS}, and then adapt  it to the planar and rotated surface code settings in 
Section~\ref{prPS}.

\subsubsection{The toric code} \label{tccPS}
Consider first the decorated graph \(G_D\) for the $[[2L^2,2,L]]$ toric code.
Recall the standard construction of the toric code from the square lattice 
by identifying the vertices and edges on the top and bottom sides,
and similarly identifying those on the left and right sides.
Let \(U\) be the set of vertices on the upper side and the left side.
These vertices are shown in orange and green in Figure~\ref{toric}. 
Let \(\tilde{G}\) be a graph obtained from \(G_D\) by deleting all vertices belonging to the gadget vertices in \(U\), together with their incident edges. 
Thus, \(\tilde{G}\)  is an \((L-1)\times (L-1)\) square-lattice graph
on the gadget vertices.

This graph admits a natural recursive planar separator into four quadrants.
We exploit this explicit partition rather than invoking the general planar
separator theorem of Lipton and Tarjan for arbitrary planar graphs,
in line with earlier separator constructions for grid graphs due to George~\cite{George73},
which predate the work of Lipton and Tarjan.

\begin{figure}[H]
  \centering
    \includegraphics[width=0.6\textwidth]{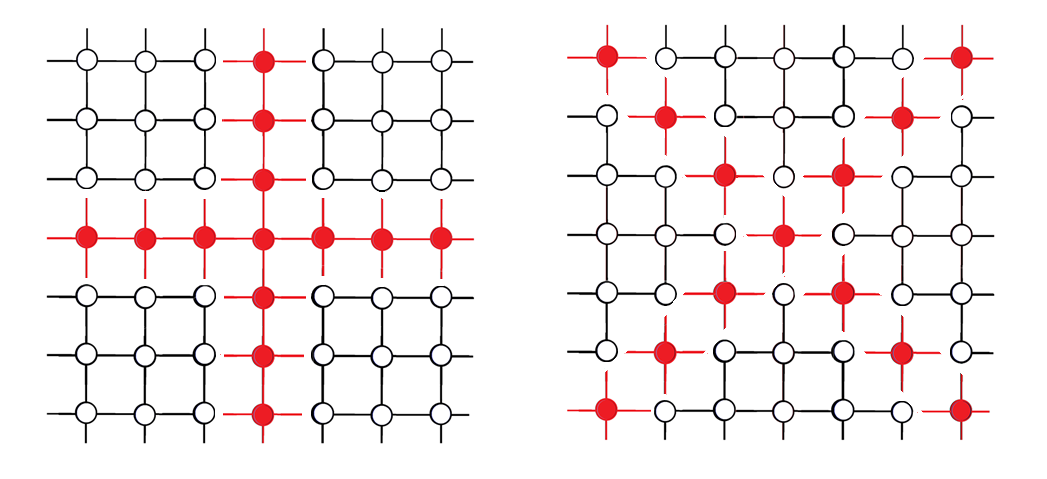}
\caption{
The \textbf{left} panel shows a portion of the trimmed decorated graph \(\tilde{G}\) associated with the toric code or planar surface code and a given set of syndrome defects.
Each circle represents a gadget vertex.
The red filled circles correspond to the gadget vertices in the separator set \(C\), consisting of gadget vertices lying on the central horizontal and vertical lines.
The black circles correspond to the gadget vertices in the four quadrants \(Q_1, Q_2, Q_3\), and \(Q_4\).
The black edges represent the edges connecting gadgets within the subgraphs \(G_1, G_2, G_3\), and \(G_4\) associated with the four quadrants. 
The \textbf{right} panel shows the separator set \(C\) and the four quadrants for the rotated surface code.
}
\label{separator}
\end{figure}

	The graph \(\tilde{G}\)  naturally decomposes into four quadrants obtained by splitting the
	\((L-1)\times (L-1)\) square lattice along its central horizontal and vertical lines.
	Let \(C\) denote the set of gadget vertices lying on the central horizontal and vertical lines, and let
	\(Q_1, Q_2, Q_3, Q_4\) denote the sets of gadget vertices in the resulting four quadrants (in arbitrary order), as illustrated in Figure~\ref{separator} on the left.
	For \(i = 1,2,3,4\), let \(G_i\) be the subgraph induced by \(\tilde{G}\)  on the  atomic vertices  contained in the gadget vertices in \(Q_i\).
	The key observation is that there are no edges in \(G_D\) connecting a vertex in \(G_i\) to a vertex in \(G_j\) for any \(i \neq j\).
	Consequently, the MWM problem on \(G_D\) can be solved recursively and independently on each \(G_i\).
	The solutions to these subproblems are then combined by applying Lemma~\ref{augPath} for  each atomic vertex  in  a gadget vertex in the  set \(C\).  
After the completion of the first recursive call, we must also
	 reintegrate the vertices of \(U\), which were removed to in order to transform the toroidal graph into a square lattice.
	This is done by applying Lemma~\ref{augPath} for each atomic vertex in each gadget of \(U\).

	Putting everything together, we obtain the following $Z$-error SMW decoder for the toric code.
	The corresponding $X$-error decoder proceeds in an analogous way on the dual lattice.

\begin{algorithm}[Toric code $Z$-error decoder]\label{toricDecoder}
\normalfont
~~\\
{\em Input:} $X$-syndrome $s_X$

\begin{itemize}
\item[1.] Construct, from the support $D$ of $s_X$ and the graph $G$ of the toric code primal lattice, the decorated graph $G_D = (V,E)$ as explained in Section~\ref{ToricReduction}.  
Assign weight $|V|-1$ to all edges of $G_D$ that connect distinct gadgets, and weight $|V|$ to all edges internal to gadgets.

\item[2.] Let $U$ be the set of vertices on the upper side and the left side of the $L\times L$ lattice, and 	
 \(\tilde{G}\) the  graph obtained from \(G_D\) by deleting all vertices belonging to the gadget vertices in \(U\), together with their incident edges.

\item[3.] Call the recursive MWM solver on \(\tilde{G}\)  to compute a maximum weight matching $M$.

\item[4.] For each gadget vertex $g$ in $U$ and each atomic vertex $v$ in $g$:
  \begin{itemize}
  \item[$\bullet$] Add $v$ to $\tilde{G}$ together with all edges 
in $G_D$  incident to $v$.
  \item[$\bullet$]  If $\tilde{G}$ admits an augmenting path with respect to $M$ having $v$ as one endpoint, find such a path $P$ of maximum net weight and update $M$ to $M \oplus P$.
  \end{itemize}

\item[5.] Return the $Z$-error corresponding to the edges of $M$ that are not internal to gadgets.
\end{itemize}

\medskip

\noindent
\textbf{Recursive MWM Solver}
~

\noindent
{\em Input:} A square $w \times w$ lattice  graph $\tilde{G}$ whose vertices are gadgets.

\begin{itemize}
\item[1.] If $w =0$ or $1$, handle the base case appropriately.

\item[2.] Let $C$ denote the set of gadget vertices lying on the central horizontal and vertical lines, and let
$Q_1, Q_2, Q_3, Q_4$ denote the sets of gadget vertices in the resulting four quadrants.

\item[3.] For $i = 1,2,3,4$,  
let \(G_i\) be the subgraph of \(\tilde{G}\) induced by the atomic vertices belonging to gadget vertices in \(Q_i\).

\item[4.] Recursively compute a maximum weight matching $M_i$ in $G_i$ for $i = 1,2,3,4$.

\item[5.] Initialize 
$M$ to $M_1 \cup M_2 \cup M_3 \cup M_4$ and 
$G'$ to $G_1 \cup G_2 \cup G_3 \cup G_4$.

\item[6.] For each gadget vertex $g$ in $C$ and each atomic vertex  $v$ in $g$:
  \begin{itemize}
  \item[$\bullet$]  Add $v$ to $G'$ together with all edges 
in $G'$  incident to $v$.
  \item[$\bullet$]  If $G'$ admits an augmenting path with respect to $M$ having $v$ as one endpoint, find such a path $P$ of maximum net weight and update
  $M$ to  $M \oplus P$.
  \end{itemize}

\item[7.] Return $M$.
\end{itemize}
\end{algorithm}

The total decoding time is therefore $O\bigl(|U|\,|V|\log |E| + T(L-1)\bigr)$, 
where the first term accounts for the total augmentation time in  Step~4 in the main algorithm, and
\(T(w)\) denotes the running time of the divide-and-conquer algorithm on a \(w \times w\) planar lattice whose vertices are gadgets.
The recurrence relation for \(T(w)\) is $$T(w) = 4T(w/2) + O(|C|\,w^2\log w),$$
where \(O(w^2 \log w)\) accounts for the time of each augmentation in Step~6 of the recursive MWM solver.
Solving this recurrence yields
$T(w) = O(|C|\,w^2\log w) = O(w^3\log w)$,
since \(|C| = \Theta(w)\).
It follows that the total decoding time is
$O\bigl(|U|\,|V|\log |E| + T(L-1)\bigr) = O(L^3\log L) = O(n^{3/2}\log n)$,  
as  \(|U| = \Theta(L)\), \(|V| = \Theta(L^2)\), and \(|E| = \Theta(L^2)\), and the number of qubits $n = 2L^2$.

In summary, we have established the following:

\begin{theorem}\label{toricSMW}
The SMW decoding problem for the toric code on $n$ qubits 
can be solved in $O(n^{3/2}\log n)$ time.
\end{theorem}

\subsubsection{Planar and  rotated surface codes 
}\label{prPS}

For the planar and rotated surface codes, the decoder is similar, except that we must handle the large planar
Fisher gadget associated with the boundary vertices (the vertices of the green triangles in Figures~\ref{decoratedPlanar} and~\ref{decoratedRotated}).

One approach is to first construct the trimmed graph
\(\tilde{G}\) by removing from the decorated graph \(G_D\) all vertices in this boundary gadget, and then operate on
\(\tilde{G}\) using the recursive MWM decoder by dividing the graph into four quadrants at the gadget level, as in the toric code case. 
After completing the recursive calls, we reinsert the removed gadget vertices through iterative augmentations,
analogous to Step~4 of Algorithm~\ref{toricDecoder}.
Unlike in the toric code case, the initial trimmed graph $\tilde{G}$ is not a perfect square lattice of gadgets; rather, it is a rectangular lattice whose side lengths may differ by a constant offset.  
The same mild rectangular asymmetry appears in the quadrants.
In the rotated surface code, the quadrants are additionally rotated, as shown in Figure~\ref{separator} (right).

Instead of deleting all vertices in the boundary gadget,
a more natural approach  is to delete only one internal vertex in the middle of the gadget. 
This turns the graph into a rectangular lattice of gadget vertices, except for special structure along two of its boundaries.
We then proceed with the recursive MWM decoder by dividing into four quadrants, while suitably handling the gadget vertices on the boundary.
After the recursion completes, we reinsert the deleted vertex  via a single augmentation step.

The only differences in the planar and rotated cases are boundary effects
(rectangular rather than square gadget lattices, and rotated quadrants in the rotated code), 
which do not affect the separator structure or the divide-and-conquer recursion.
Therefore we omit the routine details.
In either approach, the total running time is \(O(n^{3/2}\log n)\).

 Finally, recall that, as noted earlier, if we replace the nonplanar degree-4 gadgets with planar Fisher gadgets, the resulting decorated graph \(G_D\) becomes planar for both the planar and rotated surface codes.  
In a  breakthrough result~\cite{AV18}, Anari and Vazirani showed that the MWPM problem on planar graphs lies in \(\mathrm{NC}\).  
This immediately implies that the SMW decoding problem for the planar and rotated surface codes is also in \(\mathrm{NC}\), 
 and therefore 
is highly parallelizable in the sense that it admits polylogarithmic-depth circuits.

In summary, we have established the following:

\begin{theorem}\label{planarSMW}
The SMW decoding problem for the planar and rotated surface codes on \(n\) qubits 
can be solved in \(O(n^{3/2}\log n)\) time.  
Moreover, the problem lies in \(\mathrm{NC}\).
\end{theorem}

\section{FKT-based SMLC decoding} \label{FKT-SMLC}

We treat the planar and rotated surface codes and the toric code separately in
Sections~\ref{FKT-SMLCPlanar} and~\ref{tcc}, respectively.

\subsection{Planar and rotated surface codes}\label{FKT-SMLCPlanar}

In this section we give an exact SMLC decoder for the planar and rotated surface codes under independent
$X/Z$ noise modeled by two binary symmetric channels with crossover probability $p$. 
We focus first on the \(Z\)-error decoding problem; the \(X\)-error decoder is obtained analogously on the dual lattice.
Given an \(X\)-syndrome \(s_X\), represented as a primal relative \(0\)-chain,
we would like to find a \(Z\)-error \(e_Z\), represented as a primal relative
\(1\)-chain whose boundary is \(s_X\), such that the
probability
of the  coset \(e_Z + C_Z^\perp\) is maximized under the binary symmetric channel with
crossover probability \(p\).

Following Bravyi \emph{et al.}~\cite{MLCSurface},
since these codes encode one logical qubit, there are only two candidate cosets, represented by \(e_Z\) and \(e_Z+a\), where
\(e_Z\) is any \(Z\)-error whose relative boundary is \(s_X\),
and \(a\) is a relative non-contractible cycle (e.g., the minimum-length relative  
 cycle shown in green in Figures~\ref{planar} and~\ref{rotated}).
As it is easy to find such an \(e_Z\), the problem reduces to computing
the coset probability \(\pi_p(e_Z)\) for a given \(e_Z\).
The decoder outputs \(e_Z\) or \(e_Z + a\), depending on which coset has the
larger probability.
From this point forward we deviate from~\cite{MLCSurface}.

A central structural component for our SMLC decoder
is a dual formulation of coset probabilities:
Lemma~\ref{srufaceSMLCDual} shows that \(\pi_p(e_Z)\) can
be expressed as a weighted sum over all relative dual cycles. 
This duality 
is motivated by 
 the connection between toric and surface code
decoding and an Ising-type statistical mechanics model on the Nishimori line,
identified by Dennis \emph{et al.}~\cite{Dennis2002} building on the formulation
introduced by 
Nishimori~\cite{Nishimori81}. The resulting representation admits the classical
Kramers--Wannier high-temperature graphical expansion
into weighted even subgraphs
~\cite{KramersWannier1941,KramersWannier1941PartII}. 
While we were originally led to Lemma~\ref{srufaceSMLCDual} by
statistical-physics considerations, we establish it in
Section~\ref{emma} via a more general statement with a simpler proof,
based on MacWilliams duality and Fourier analysis.
The advantage of this alternative proof is that it is purely algebraic and applies
to arbitrary linear codes; the geometric interpretation for surface and toric codes
enters only through the geometric definition of the code \(C_Z\) in the CSS construction. 
This duality also admits a natural extension to the depolarizing channel,
as explained in Section~\ref{depolS}.

Once \(\pi_p(e_Z)\) is expressed as a sum over dual cycles, we use Fisher gadgets in their original role~\cite{Fisher66}
to convert the cycle sum into a weighted sum over
perfect matchings on a decorated \emph{planar} graph. 
We then invoke Kasteleyn’s Pfaffian method~\cite{Kasteleyn1963} to evaluate this planar
perfect-matching sum via a Pfaffian, equivalently via the determinant of a
skew-symmetric matrix.
This Fisher--Kasteleyn pipeline is the standard mechanism behind what is commonly
referred to as the (Fisher--Kasteleyn--Temperley) FKT/Pfaffian method for counting
perfect matchings in a planar graph.

In what follows, we discuss each of the above steps in detail.

\begin{lemma}\label{srufaceSMLCDual}
\textnormal{\bf(Kramers--Wannier--MacWilliams duality for planar and rotated surface codes)} 
 Consider the  planar surface code or the rotated surface code  as defined in Sections~\ref{planarS} and \ref{rotatedS}, with
\(\mathcal{X}\) and \(\mathcal{A}\)  
denoting the underlying primal cell complex  and its 
1-dimensional
subcomplex, \(\mathcal{X}^\ast\) and \(\mathcal{A}^\ast\) denoting their duals, 
and
\(C_X\) and \(C_Z\) the classical codes underlying the CSS construction of these codes. 

Let \(e_Z\) be a primal error represented as a relative \(1\)-chain.
Let \(\pi_p(e_Z)\) denote the probability of the coset \(e_Z + C_Z^\perp\) under the binary symmetric channel with crossover probability \(p\). Then
\[
\pi_p(e_Z) = 2^{-(n-m)}
\sum_{c\ \text{relative dual cycle}}~~
\prod_{e \in c}
w^{(e_Z)}_e,
\]
where \(m\) is the number of  plaquettes in $\mathcal{X}$, 
 \(n\) is the number of qubits, and
\[
w^{(e_Z)}_e = (-1)^{(e_Z)_{e^\ast}} (1 - 2p),
\]
with \(e^\ast\) denoting the primal edge associated with the dual edge \(e\).
\end{lemma}
Now we use Fisher gadgets in their original role, as introduced by Fisher~\cite{Fisher66},
to convert a sum over cycles into a sum over perfect matchings.

\begin{figure}[H]
  \centering
    \includegraphics[width=0.65\textwidth]{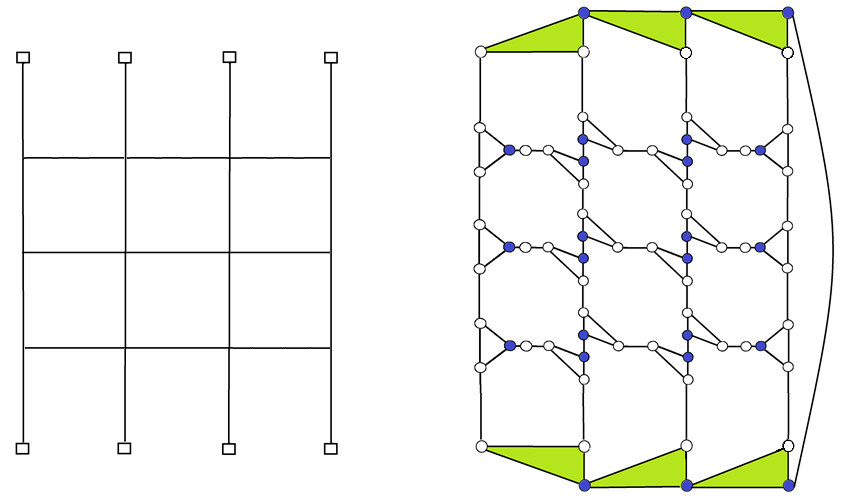}
\caption{  
{\bf Left:} The graph \(G^\ast\) for the planar surface code for $L=4$, with the vertices in \(B^\ast\) shown as boxes.
{\bf Right:} The planar decorated graph \(G^\ast_{\emptyset}\). 
}
\label{decoratedPlanarSMLC}
\end{figure}

\begin{figure}[H]
  \centering
    \includegraphics[width=0.9\textwidth]{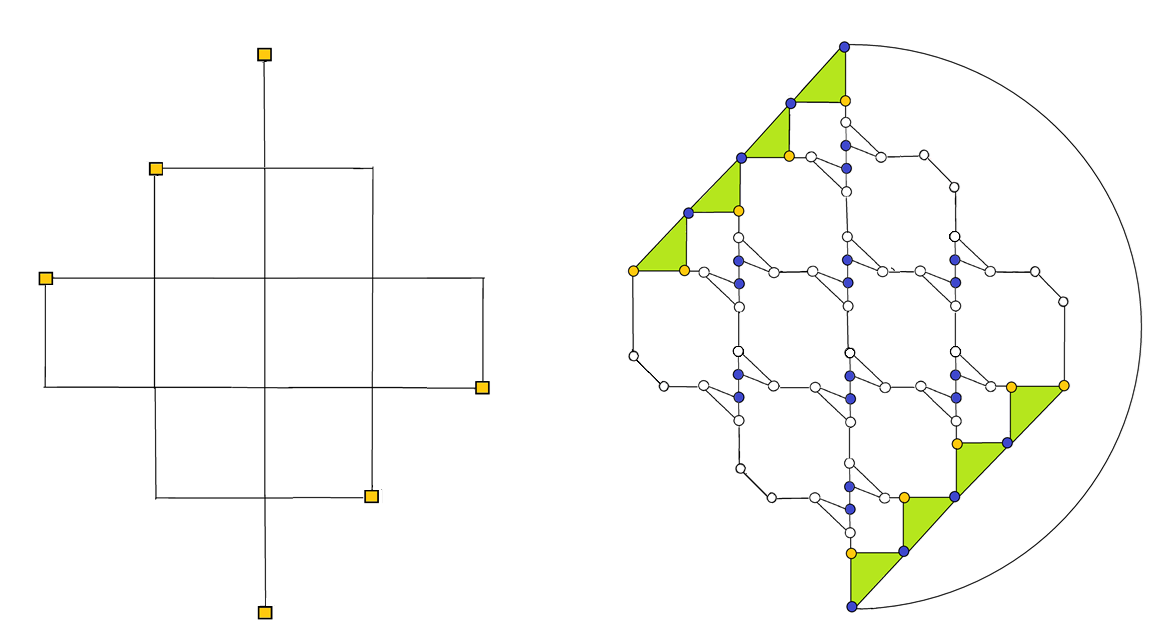}
\caption{  
{\bf Left:} The graph \(G^\ast\) for the rotated surface code for $L=5$, with the vertices in \(B^\ast\) shown as boxes.
{\bf Right:} The planar decorated graph \(G^\ast_{\emptyset}\). 
}
\label{decoratedRotatedSMLC}
\end{figure}

Construct the decorated graph \(G^\ast_\emptyset\) from the dual complex
\(\mathcal{X}^\ast\) and its \(1\)-dimensional subcomplex \(\mathcal{A}^\ast\),
as explained in Sections~\ref{redPlanar} and~\ref{redRot} with respect to the
empty defect set.
Here we use planar degree-$4$ Fisher gadgets to ensure that the decorated graph
is planar.
The planarity of the graph is essential for the Pfaffian reduction that follows.
See Algorithm~\ref{SMLCDec} for a self-contained description of the construction, and
Figures~\ref{decoratedPlanarSMLC} and~\ref{decoratedRotatedSMLC}
for illustrations.

By construction, there is a one-to-one correspondence between relative dual cycles
and perfect matchings in the decorated graph \(G^\ast_\emptyset\).
To map the cycle weight to a matching weight, extend the weights
\(w^{(e_Z)}_e\) defined in Lemma~\ref{srufaceSMLCDual} by setting the weight of
each gadget-internal edge \(e\) to \(w^{(e_Z)}_e = 1\). Let \(S_{e_Z}=2^{n-m}\pi_p(e_Z)\). Then  
\[
S_{e_Z} = \sum_{c\ \text{relative dual cycle}}~~
\prod_{e \in c}
w^{(e_Z)}_e
=
\sum_{M \in \mathcal{M}_{G^\ast_\emptyset}}~~
\prod_{e \in M} w^{(e_Z)}_e,
\]
where \(\mathcal{M}_{G^\ast_\emptyset}\) denotes the set of perfect matchings of
\(G^\ast_\emptyset\).
Note that, unlike the MWPM setup—where the weight of a matching is computed
by summing the weights of its edges—here the matching weight is given by the
product of its edge weights.

Now we invoke Kasteleyn’s Pfaffian method for counting weighted perfect
matchings in planar graphs~\cite{Kasteleyn1963}; we recall the necessary
background on Pfaffians and Pfaffian orientations in
Appendix~\ref{app:pfaffians}.
Compute a Pfaffian orientation \(\vec{E}\) of \(G^\ast_\emptyset\).
Thus
\[
S_{e_Z}= \pm \operatorname{Pf}(A^{(e_Z)}),
\]
where the skew-symmetric matrix \(A\) is defined by
\begin{equation}\label{matrixEnc}
A^{(e_Z)}_{u,v} =
\begin{cases}
\phantom{-}w^{(e_Z)}_{(u,v)} &\mbox{if } (u,v) \in \vec{E},\\
-w^{(e_Z)}_{(u,v)} &\mbox{if }  (v,u) \in \vec{E},\\
0 & \text{otherwise}.
\end{cases}
\end{equation}
Since \(\operatorname{Pf}(A^{(e_Z)})^2 = \det(A^{(e_Z)})\)  
 and \(S_{e_Z}  \ge 0\), as
\(2^{-(n-m)} S_{e_Z}  = \pi_p(e_Z)\) is  a probability,
we obtain \(S_{e_Z}  = \sqrt{\det(A^{(e_Z)})}\).

Putting everything together, we obtain the following $Z$-error  decoder.
The corresponding $X$-error decoder proceeds in an analogous way on the dual lattice.

\begin{algorithm}[SMLC \(Z\)-error decoder
for planar and rotated surface codes]\label{SMLCDec}
\normalfont
~~\\
{\em Input:}  an \(X\)-syndrome \(s_X\), represented as a primal relative \(0\)-chain.
\begin{itemize}
\item[1.] 
Find any \(Z\)-error \(e_Z\) whose relative boundary 
is $s_X$

\item[2.] 
Let \(G^\ast\) be the graph of \(\mathcal{X}^\ast \setminus \mathcal{A}^\ast\), obtained by removing all edges of
\(\mathcal{A}^\ast\) and deleting any vertex of \(\mathcal{A}^\ast\) that becomes isolated, and let 
 \(B^\ast\) be the set of remaining vertices in  \(\mathcal{A}^\ast\). 
 Construct the decorated graph \(G^\ast_\emptyset\) from $G^\ast$ as follows. 
Replace each vertex of \(G^\ast \setminus B^\ast\) by an even Fisher gadget in \(G^\ast_\emptyset\) of degree \(4\) or degree \(2\), as appropriate. 
Connect gadget external vertices according to the adjacency structure of \(G^\ast\).
Replace \(B^\ast\) in \(G^\ast_\emptyset\) by a single even-parity Fisher gadget
connected to external gadget
vertices associated with vertices in \(G^\ast \setminus B^\ast\) incident to
\(B^\ast\). See Figures~\ref{decoratedPlanarSMLC} and~\ref{decoratedRotatedSMLC}.

\item[3.]   
Assign weights \(w^{(e_Z)}\) to the edges of \(G^\ast_\emptyset\) as follows.
All gadget-internal edges $e$ receive weight \(w^{(e_Z)}_e = 1\).
Each edge inherited from \(G^\ast\) receives weight
\[
w^{(e_Z)}_{e} =
\begin{cases}
-(1-2p) &\mbox{if } e^\ast \in e_Z\\
\ \ 1-2p &\mbox{otherwise},
\end{cases}
\]
where $e^\ast$ is the  primal edge corresponding to $e$.

\item[4.]  Compute a Pfaffian orientation \(\vec{E}\) of \(G^\ast_\emptyset\).

\item[5.]  Compute the nonzero entries of the   
 $n\times n$ skew-symmetric matrix   $A^{(e_Z)}$ obtained  from the adjacency matrix 
of $G^\ast_\emptyset$, the orientation $\vec{E}$, and 
the weights $w^{(e_Z)}$ via (\ref{matrixEnc}). 

\item[6.] Compute  the
determinant $\det(A^{(e_Z)})$.

\item[7.] Repeat Steps 3, 5, and 6 for $e'_Z = e_Z + a$,
where $a$ is a non-contractible primal relative cycle.

\item[8.] If $\det(A^{(e_Z)}) > \det(A^{(e'_Z)})$, return the
$Z$-error $e_Z$; else return $e'_Z$.

\end{itemize}

\end{algorithm}

In what follows we analyze the running time of the decoding algorithm.
Finding an initial error \(e_Z\) in Step~1 can be done efficiently in time
linear in the number \(n\) of qubits, for instance using the peeling algorithm
of Delfosse and Z\'emor~\cite{DZ-Erasure}.
Pfaffian orientations in Step~4 can also be computed in linear time for planar
graphs~\cite{KarpinskiRytter98}.

Since the matrix \(A^{(e_Z)}\) constructed in Step~5 is supported on the planar
graph \(G^\ast_\emptyset\), its determinant can be computed asymptotically faster
than the general \(O(n^{\omega})\) bound on the number of arithmetic operations. 
In particular, Yuster~\cite{Yuster} showed that the absolute value of the determinant of an
\(n \times n\) matrix supported by a planar graph can be computed in
\(O(n^{\omega/2})\) arithmetic operations.
This result builds on the seminal nested dissection method of
Lipton, Rose, and Tarjan~\cite{LiptonRoseTarjan79},  
and on the planar separator framework of Lipton and Tarjan~\cite{LiptonTarjan79},
which in turn extends earlier work of George on planar grids~\cite{George73}. 
Thus, the efficiency of both the SMLC decoder and the SMW decoder
in Section~\ref{planarSep} ultimately relies on classical separator-based ideas,
originating in \cite{George73,LiptonTarjan79}.

The fact that the algorithm in~\cite{Yuster} computes the absolute value of the
determinant is not restrictive in our setting, since the matrix
\(A^{(e_Z)}\) is real and skew-symmetric of even order, and hence satisfies
\(\det(A^{(e_Z)}) \ge 0\), as
\(\operatorname{Pf}(A^{(e_Z)})^2 = \det(A^{(e_Z)})\).

The \(O(n^{\omega/2})\) bound refers to algebraic complexity.
Assuming that the matrix entries admit \(O(1)\)-bit representations,
the corresponding bit complexity is \(\tilde{O}(n^{\omega/2+1})\).
This assumption applies here: the edge weights are \(1\) or
\(\pm(1-2p)\).
Assuming that \(p\) is a rational number with an \(O(1)\)-bit representation,
the computation can be made exact by clearing denominators after rescaling.

Finally, since determinant computation lies in \(\mathrm{NC}\)~\cite{Csa76},
the SMLC decoding problem for planar and rotated surface codes
lies in \(\mathrm{NC}\) as well
\footnote{ 
To see why the other steps of Algorithm~\ref{SMLCDec} are computable in $\mathrm{NC}$, first note that the computation of a Pfaffian orientation is in $\mathrm{NC}$~\cite{Vaz89}. 
Since  the SMW decoding problem for the planar and rotated surface codes is in $\mathrm{NC}$,
it follows that the simpler task of computing an initial error $e_Z$ in Step~1
is also in $\mathrm{NC}$. 
A more direct $\mathrm{NC}$ algorithm for finding $e_Z$ proceeds as follows:
pair the defects in parallel according to a prespecified ordering of the vertices
of $G^\ast$, pairing the last and first elements of $B^\ast$ if the number of
defects is odd. Then, for each resulting pair, generate in parallel a Manhattan-type
path, and finally output the XOR of all these paths.}.

In summary, we have established the following:

\begin{theorem}\label{planarRotSMLC}
The SMLC decoding problem for the planar and rotated surface codes on \(n\) qubits
can be solved in \(O(n^{\omega/2})\) arithmetic operations and
\(\tilde{O}(n^{\omega/2+1})\) bit operations.
Moreover, the problem lies in \(\mathrm{NC}\).
\end{theorem}

\subsubsection{Proof of Lemma \ref{srufaceSMLCDual}}\label{emma}

Lemma~\ref{srufaceSMLCDual} follows by 
specializing Lemma~\ref{genlem} below to planar and rotated surface codes, and noting 
 that for these codes 
 \(C_Z\) consists of relative dual cycles and has size \(2^{n-m}\), since the
\(m\) plaquette \(Z\)-checks of \(\mathcal{X}\) are linearly independent.

The proof of Lemma~\ref{genlem} uses elementary notions from 
Fourier analysis on the hypercube. We start with some preliminaries. 
The study of error-correcting codes via harmonic analysis dates back to
MacWilliams~\cite{Mac63}; see also~\cite{LMN93} and references therein.
Identify the hypercube \(\{0,1\}^n\) with the abelian group
\(\mathbb{Z}_2^n = (\mathbb{Z}/2\mathbb{Z})^n\).
The characters of \(\mathbb{Z}_2^n\) are the functions
\(\{\chi_z\}_{z \in \mathbb{Z}_2^n}\),
where \(\chi_z : \{0,1\}^n \to \{\pm1\}\) is given by
$\chi_z(x) = (-1)^{\langle x, z \rangle}$ and 
$\langle x, z \rangle = \sum_{i=1}^n x_i z_i$. 
Let \(\mathcal{L}(\mathbb{Z}_2^n)\) denote the vector space of complex-valued functions
on \(\mathbb{Z}_2^n\), equipped with the inner product
\[
\langle f,g\rangle
=
\mathbb{E}[f \overline{g}]
=
\frac{1}{2^n}\sum_{x \in \mathbb{Z}_2^n} f(x)\overline{g(x)}.
\]
The {\em characters} \(\{\chi_z\}_z\) form an orthonormal basis of
\(\mathcal{L}(\mathbb{Z}_2^n)\).
For \(f \in \mathcal{L}(\mathbb{Z}_2^n)\), its {\em Fourier transform} 
\(\widehat{f}\) is defined by
\[
f(x) = \sum_{z} \widehat{f}(z)\chi_z(x) 
~~\mbox{and}~~ 
\widehat{f}(z) = \langle f ,\chi_z\rangle =
\mathbb{E}[f \chi_z].
\]
We will use the following standard facts.
\begin{fact}[Fourier transform of the  exponential function]\label{ffacts1}
If 
$g_r:\{0,1\}^n\rightarrow \CN$  is given by $g_r(x) = r^{|x|}$, where 
$r$ is a complex number,  
then  
$\widehat{g_r}(z) =\left( \frac{1+r}{2}\right)^n \left(\frac{1-r}{1+r}\right)^{|z|}$. 
\end{fact}

\begin{fact}[Fourier transform and linear codes]\label{ffacts2}
If $C\subset \F_2^n$  is an $\F_2$-linear code and $C^\perp$ is its dual, then for each $z\in \F_2^n$, we have 
\[
  \sum_{y\in C^\perp} \chi_z(y)    =\left\{\begin{array}{ll} |C^\perp| & \mbox{ if  } z\in C\\ 0 & \mbox{ otherwise. }\end{array}\right.
\]
\end{fact}
See for instance \cite{Baz15} for  proofs of Facts \ref{ffacts1} and \ref{ffacts2}.
\begin{lemma}[Dual-code coset probabilities]\label{genlem}
Let \(C \subset \mathbb{F}_2^n\) be a linear code.
Let \(e \in \mathbb{F}_2^n\), and let \(\pi_p(e)\) denote the probability of the coset
\(e + C^\perp\) under the binary symmetric channel with crossover probability \(p\).
Then
\[
\pi_p(e)
=
\frac{1}{|C|}
\sum_{z \in C}
\prod_{i \in z} w_i^{(e)},
\]
where 
$w_i^{(e)} = (-1)^{e_i}(1-2p)$.
\end{lemma}
\emph{Proof.}
We have
\[
\pi_p(e)
=
\sum_{y \in C^\perp} (1-p)^{n-|e+y|} p^{|e+y|}
=
(1-p)^n \sum_{y \in C^\perp} g_r(e+y),
\]
where \(r = p/(1-p)\).
Expanding \(g_r\) in the Fourier basis,
\[
g_r(x) = \sum_z \widehat{g_r}(z)\chi_z(x),
\]
and using \(\chi_z(e+y)=\chi_z(e)\chi_z(y)\), we obtain
\[
\pi_p(e)
=
(1-p)^n
\sum_z \widehat{g_r}(z)\chi_z(e)
\sum_{y \in C^\perp} \chi_z(y).
\]
By Fact~\ref{ffacts2}, only terms with \(z \in C\) contribute, giving
\[
\pi_p(e)  =
(1-p)^n|C^\perp|
 \sum_{z\in C} \widehat{g_r}(z) \chi_z(e) 
 = |C^\perp|\left( \frac{(1-p)(1+r)}{2}\right)^n
 \sum_{z\in C}
 \left(\frac{1-r}{1+r}\right)^{|z|}\chi_z(e), 
\]
where the second equality uses  Fact \ref{ffacts1}.
The lemma then follows from the identities 
$(1-p)(1+r) = 1$, $\frac{1-r}{1+r}=1-2p$, 
and  \(|C^\perp|=2^{n}/|C|\).
\finito

\subsection{The toric code}\label{tcc}

In this section we extend the SMLC decoder to the toric code.
The planarity of the planar and rotated surface codes
removes the Pfaffian sign ambiguity in our application.
For the toric code, the analogous reduction lives on a genus-\(1\) surface, and resolving the Pfaffian sign is more costly.

As before, we focus first on the \(Z\)-error decoding problem; the \(X\)-error decoder is obtained analogously on the dual lattice.
Let \(\mathcal{X}\) and \(\mathcal{X}^{\ast}\) be the primal and dual lattices of the toric code.
Given an \(X\)-syndrome \(s_X\), represented as a primal \(0\)-chain,
we would like to find a \(Z\)-error \(e_Z\), represented as a primal
\(1\)-chain whose boundary is \(s_X\), such that the stabilizer coset probability
of \(e_Z + C_Z^\perp\) is maximized under the binary symmetric channel with
crossover probability \(p\).

Since the code encodes two logical qubits, we have
four candidate cosets, represented by \(e_Z\), \(e_Z+a\),
\(e_Z+b\), and \(e_Z+a+b\), where
\(e_Z\) is any \(Z\)-error whose boundary is \(s_X\),
and \(a\) and \(b\) are two non-contractible cycles that are
not homotopic to each other
(for example, the minimum-length  cycles shown in yellow and green in Figure~\ref{toric}).
Thus, as before, the problem reduces to computing
the coset probability \(\pi_p(e_Z)\) for a given \(e_Z\).

The following analogue of Lemma~\ref{srufaceSMLCDual}, which follows immediately
from Lemma~\ref{genlem}, expresses \(\pi_p(e_Z)\) as a
weighted sum over all dual cycles.

\begin{lemma} \label{toricSMLCDual}
\textnormal{\bf(Kramers--Wannier--MacWilliams  duality for the  toric code)}
Consider the toric code as defined in Section~\ref{toricS},
with \(\mathcal{X}\) and \(\mathcal{X}^\ast\) denoting the underlying primal cell complex and its dual, and
\(C_X\) and \(C_Z\) the classical codes underlying the CSS construction of the toric code.

Let \(e_Z\) be a primal error represented as a \(1\)-chain.
Let \(\pi_p(e_Z)\) denote the probability of the coset \(e_Z + C_Z^\perp\) under the binary symmetric channel with crossover probability \(p\).
Then
\begin{equation}
\pi_p(e_Z) = 2^{-(n-m+1)}
\sum_{c\ \text{dual cycle}}~~
\prod_{e \in c}
w^{(e_Z)}_{e},
\end{equation}
where \(m\) is the number of plaquettes in \(\mathcal{X}\), \(n\) is the number of qubits, and
\[
w^{(e_Z)}_{e} = (-1)^{(e_Z)_{e^\ast}} (1 - 2p),
\]
with \(e^\ast\) denoting the primal edge associated with the dual edge \(e\).
\end{lemma}
Lemma~\ref{toricSMLCDual} follows by specializing Lemma~\ref{genlem} to the toric code and noting
that for this code \(C_Z\) consists of dual cycles and has size \(2^{n-m-1}\),
since the \(m\) plaquette \(Z\)-checks of \(\mathcal{X}\) are not linearly independent;
namely, their sum is zero because the torus is a closed surface.

Now, as before, to turn the cycle sum into a sum over perfect matchings, we construct
the decorated graph \(G^\ast_\emptyset\) from the dual lattice \(\mathcal{X}^\ast\).
The construction is much simpler than in the surface-code case, as it only uses
degree-\(4\) Fisher gadgets for each vertex of \(\mathcal{X}^\ast\).
We extend the weights \(w^{(e_Z)}_e\) defined in Lemma~\ref{toricSMLCDual} by setting the weight of
each gadget-internal edge \(e\) to \(w^{(e_Z)}_e = 1\).
Thus,
\[
2^{n-m-1}\pi_p(e_Z) = \sum_{c\ \text{ dual cycle}}~~
\prod_{e \in c}
w^{(e_Z)}_e
=
\sum_{M \in \mathcal{M}_{G^\ast_\emptyset}}~~
\prod_{e \in M} w^{(e_Z)}_e,
\]
where \(\mathcal{M}_{G^\ast_\emptyset}\) is  the set of perfect matchings of \(G^\ast_\emptyset\).

Unlike the surface-code case, where \(G^\ast_\emptyset\) is planar, the graph
\(G^\ast_\emptyset\) has genus \(1\). Consequently, the weighted sum over perfect
matchings is no longer given by a single Pfaffian up to sign.
Galluccio and Loebl~\cite{GalluccioLoebl} extended Kasteleyn’s work to genus-\(g\)
graphs by showing that, for such graphs, the weighted sum over all perfect
matchings can be expressed as a linear combination of \(4^{g}\) Pfaffians with
appropriate sign choices.
Below we state their result specialized to the genus-\(1\) case.

\begin{theorem}[Galluccio--Loebl, 1999~\cite{GalluccioLoebl}]\label{4paff}
Let \(G = (V,E)\) be a genus-\(1\) graph that is \(2\)-connected and admits a perfect
matching.
Then there exist four efficiently computable edge orientations
\(\vec{E}_{1,1}\), \(\vec{E}_{-1,1}\), \(\vec{E}_{1,-1}\), and \(\vec{E}_{-1,-1}\) of \(G\)
such that for any fixed perfect matching \(M_0\) and any edge-weight function
\(w : E \to \mathbb{R}\),
\begin{eqnarray*}
2\sum_{M\in \mathcal{M}_G}
\prod_{e\in M} w_e &=& 
\mathrm{sgn}_{\vec{E}_{1,1}}(M_0)\operatorname{Pf}(A^{(1,1)})
+
\mathrm{sgn}_{\vec{E}_{-1,1}}(M_0)\operatorname{Pf}(A^{(-1,1)})
+
\mathrm{sgn}_{\vec{E}_{1,-1}}(M_0)\operatorname{Pf}(A^{(1,-1)})\\
&&\quad-
\mathrm{sgn}_{\vec{E}_{-1,-1}}(M_0)\operatorname{Pf}(A^{(-1,-1)}),
\end{eqnarray*}
where \(A^{(r,s)}\) is the \(V \times V\) skew-symmetric matrix defined by 
$A^{(r,s)}_{i,j} = w_{i,j}$ if $(i,j)\in \vec{E}_{r,s}$ 
and $A^{(r,s)}_{i,j} =-w_{i,j}$, otherwise,  for all $r,s\in \{\pm1\}$.
\end{theorem}
In contrast to the planar case, the signs of the Pfaffians appearing above are no
longer irrelevant, since we would like to compute a sum of four Pfaffians.  
As a result, computing only the absolute value of the determinant of
\(A^{(r,s)}\) and using the identity
 \(\operatorname{Pf}(A^{(r,s)})^2 = \det(A^{(r,s)})\) 
is insufficient. Accordingly, 
while there exist efficient algorithms for computing the absolute value of the
determinant for bounded-genus graphs in
\(O(n^{\omega/2})\) arithmetic operations~\cite{AlonYuster}, these algorithms do not apply here
because the sign of the Pfaffian is also required.
The Pfaffian together with its sign can be computed in
\(O(n^{3})\) arithmetic operations~\cite{GalbiatiMaffioli}.
In contrast to the determinant, it is not known whether the problem of computing
the Pfaffian lies in \(\mathsf{NC}\).

We summarize the above discussion in the following result.

\begin{theorem}\label{troicSMLC4}
The SMLC decoding problem for the toric code on \(n\) qubits
can be solved in \(O(n^{3})\) arithmetic operations and
\(\tilde{O}(n^{4})\) bit operations.
\end{theorem}
Although the resulting complexity is higher than in the planar case, the above theorem
establishes that exact SMLC decoding on a genus-$1$ surface is nevertheless
tractable in polynomial time.

\section{Depolarizing channel}\label{depolS}

While the independent \(X\)- and \(Z\)-noise model admits an exact decomposition
into two independent decoding problems, physically relevant noise models are
typically better approximated by the depolarizing channel, under which \(X\)- and
\(Z\)-errors are correlated.
Specifically, each qubit is left unchanged with probability \(1-p\), and undergoes
an \(X\), \(Y\), or \(Z\) error with probability \(p/3\) each.

For a CSS code under the depolarizing channel, given an \(X\)-syndrome \(s_X\) and
a \(Z\)-syndrome \(s_Z\), \emph{Minimum Weight (MW) decoding} selects a pair of
errors \(e_Z\) and \(e_X\), consistent respectively with \(s_X\) and \(s_Z\), that
maximizes the probability of a single error representative \((e_Z,e_X)\).
Equivalently, this amounts to minimizing the Hamming weight of
\(e_Z \vee e_X\), since a qubit supports a \(Y\)-error precisely when both
\(Z\)- and \(X\)-components are present, where \(\vee\) denotes the bitwise OR.

The MW decoding objective  ignores stabilizer degeneracy.  
In contrast, given a syndrome pair \((s_X,s_Z)\),
\emph{Most Likely Coset (MLC) decoding} selects the stabilizer coset
\((e_Z,e_X) + C_Z^\perp \times C_X^\perp\) whose total probability under the
depolarizing channel is maximal, where the errors \(e_Z\) and \(e_X\) are
consistent respectively with the syndromes \(s_X\) and \(s_Z\).

Several approximate approaches to joint decoding under depolarizing noise have
been proposed.
Methods that are partially aware of noise correlations are introduced
in~\cite{DTcor,YuanLu2022} in the context of MW decoding of surface codes.
In~\cite{MLCSurface}, the MLC decoding problem for the planar surface code is
formulated as a tensor-network contraction.
Using a canonical tensor-network approximation algorithm, the authors obtain an
approximate decoder that runs in time \(O(n\,\chi^{3})\), where \(\chi\) is a
parameter controlling the approximation accuracy.

In~\cite{FischerMiyake}, it is shown that for surface codes the MW and MLC decoding
problems are NP-hard and \#P-hard, respectively, under non-identically distributed
depolarizing noise, where qubits undergo depolarizing errors with probabilities
that depend both on the qubit and on whether the error is of type \(X\), \(Y\), or
\(Z\).
The established hardness relies on carefully constructed, spatially varying error
probabilities.
This result does not, however, imply intractability of MW or MLC decoding under
the identically distributed depolarizing channel.
It nevertheless suggests that these problems are difficult, since
for the non-identically distributed binary symmetric channel both MW and SMLC
decoding for surface codes remain tractable (see
Section~\ref{extAndOpenQuestions}).

The results reported in this paper were originally motivated by the study of
decoding under depolarizing noise.
Consider, for instance, the toric code.
To reduce the SMW decoding problem to instances of MWPM, the classical approach
operates on a complete graph on the \(X\)-syndrome defects for correcting
\(Z\)-errors, and a separate complete graph on the \(Z\)-syndrome defects for
correcting \(X\)-errors.
While joint decoding problems require reasoning locally about the coupling between
\(X\)- and \(Z\)-errors on each qubit, this locality is entirely lost in the
complete decoding graphs.
This issue led us  to an alternative approach in which the decoding problem
is reduced to MWPM through the use of  local gadgets that preserve the
geometric structure of the code.
Although this local perspective does not presently yield a solution to the MW
decoding problem under depolarizing noise, it proved instrumental in obtaining the
improvements for the independent-noise model reported in this paper.
It is nevertheless worthwhile to formulate both the MW and MLC decoding problems
for surface codes within this local framework, which we do next, starting with the MW problem.

\subsection{Local formulation of MW decoding}

Consider the toric code.
Let \(D_X\) and \(D_Z\) denote the supports of the \(X\)- and \(Z\)-syndromes,
respectively.
Starting from the primal lattice graph \(G\), construct the decorated graph
\(G_{D_X}\) by replacing each vertex in \(D_X\) with an odd degree-$3$ Fisher
gadget, and each vertex of \(G\) not in \(D_X\) with an even degree-$4$ Fisher
gadget (or its nonplanar variant).
As before, the external vertices of the gadgets are connected according to the
adjacency structure of \(G\).
Similarly, starting from the dual lattice graph \(G^\ast\), construct the
decorated graph \(G^\ast_{D_Z}\) using the defect set \(D_Z\).

Note that, in contrast to the separate decoding setup, we cannot model each defect using a
single vertex in the decorated graphs, since we don't  know whether the degree-$3$ defect 
resolution theorem (Theorem \ref{DJoin}) continues to hold under the depolarizing channel.

Consider the primal-dual pairing between the edges of
$G=(V_X,E)$ and those of $G^\ast=(V_Z,E^\ast)$, which associates each edge
$e\in E$ with a dual edge $e^\ast\in E^\ast$.
If $M_X$ and $M_Z$ are perfect matchings of $G_{D_X}$ and $G^\ast_{D_Z}$,
respectively, define their \emph{joint weight} as
 $\sum_{e \in E} x_e$, 
where  $x_e = 1$ 
if $e \in M_X$ {\em or} $e^{\ast} \in M_Z$. 
That is, gadget-internal edges contribute zero weight, while each edge inherited
from \(G\) or \(G^\ast\) contributes weight one, with no double counting of a
primal edge and its dual.

Accordingly, the MW decoding problem reduces to the
\emph{Joint Minimum Weight Perfect Matching (JMWPM)} problem: find a perfect
matching \(M_X\) of \(G_{D_X}\) and a perfect matching \(M_Z\) of \(G^\ast_{D_Z}\) whose
joint weight is minimized.

\begin{openproblem}\label{open1}
Does the JMWPM problem admit an efficient algorithm?
\end{openproblem}
As in Sections~\ref{planarS} and~\ref{rotatedS}, the planar and rotated surface
codes can be handled analogously, using even degree-$2$ or degree-$3$ Fisher
gadgets when applicable, and a single Fisher gadget at boundary vertices, whose
parity is determined by the parity of the corresponding defect set.


\medskip
\noindent\emph{Note added.}
This question was subsequently resolved negatively in~\cite{GWK26}
and independently in~\cite{BK26}, which establish NP-hardness of
MW decoding for surface codes. 
Moreover, \cite{BK26} establishes hardness of approximation within a
polynomial additive gap, showing that, for some positive constant $c$,
no polynomial-time algorithm for MW decoding can find $X$- and $Z$-errors
consistent with the given syndromes whose joint weight is within
$c n^\alpha$ of the optimum, where $\alpha=1/14$ for the toric code
and $\alpha=1/18$ for the planar surface code, unless
$\mathrm{P}=\mathrm{NP}$. 
 On the positive side, \cite{GWKApprox26} gives a polynomial-time
approximation scheme, which, for every fixed $\epsilon>0$, returns a
solution of joint weight at most $(1+\epsilon)$ times the optimum.
It remains open whether MW decoding is hard to approximate within an
additive gap of $O(n^\alpha)$ for $1/14<\alpha<1$ for the toric code
and $1/18<\alpha<1$ for the planar surface code. 
Can the local JMWPM formulation developed here be exploited to design
efficient practical decoding schemes?

\subsection{MLC decoding}

Consider again the toric code.
Since the code encodes two logical qubits, the MLC decoding problem reduces to
computing the probabilities of the \(16\) stabilizer cosets of the form
\((e_Z,e_X) + C_Z^\perp \times C_X^\perp\), for suitably chosen and efficiently
computable errors \(e_Z\) and \(e_X\) that are consistent with the syndromes
\(s_X\) and \(s_Z\), respectively.
In the planar and rotated surface code cases, there are only \(4\) such cosets.

We show in Corollaries~\ref{depolToric} and~\ref{depolSurface} that each of these
coset probabilities can be expressed as a \emph{jointly weighted} sum over all
primal and dual cycles.
Both corollaries follow from Lemma~\ref{genlemDep}, which is the analogue of
Lemma~\ref{genlem} for the depolarizing channel.

\begin{lemma}[Dual-code coset probabilities on the depolarizing channel]\label{genlemDep} 
Consider the following channel with parameter \(p\) on
\(\mathbb{F}_2^{2n}\), where \(0<p<1\). 
For  \(i=1,\ldots,n\), the pair \((x_i,x_i')\) takes the value
\((0,0)\) with probability \(1-p\), and each of the values
\((0,1),(1,0),(1,1)\) with probability \(p/3\).
The $n$ pairs \(\{(x_i,x_i')\}_{i=1}^n\) are statistically independent.

Let \(C,C' \subset \mathbb{F}_2^n\) be linear codes and
let \(e,e' \in \mathbb{F}_2^n\).
Let \(\pi_p(e,e')\) denote the probability of the coset
\((e,e') + C^\perp \times C'^\perp\).
Then
\[
\pi_p(e,e')
=
\frac{1}{|C|\,|C'|}
\sum_{z \in C}\sum_{z' \in C'}
\chi_z(e)\chi_{z'}(e') \,
\alpha_p^{|z \vee z'|},
\]
where
$\alpha_p = 1 - \frac{4p}{3}$ 
and \(z \vee z'\) denotes the bitwise OR of \(z\) and \(z'\).
\end{lemma}
The proof of Lemma~\ref{genlemDep} is based on Fourier analysis and is given in
Section~\ref{genlemDepProof}.

\begin{corollary}\label{depolToric}
\textnormal{\bf (Kramers--Wannier--MacWilliams duality for the toric code under depolarizing noise).}
Consider the toric code as defined in Section~\ref{toricS},
with \(\mathcal{X}\) and \(\mathcal{X}^\ast\) denoting the underlying primal cell complex and its dual, and
\(C_X\) and \(C_Z\) the classical codes underlying the CSS construction of the toric code.

Let \(e_Z\) and \(e_X\) be primal and dual errors, represented respectively as a primal \(1\)-chain
and a dual \(1\)-chain.
Let \(\pi_p(e_Z,e_X)\) denote the probability of the coset
$(e_Z,e_X) + C_Z^\perp \times C_X^\perp$ 
under the depolarizing channel with parameter \(p\).
Then
\[
\pi_p(e_Z,e_X)
=
4^{-(n-m+1)}
\sum_{\substack{
       c_Z \,\text{dual cycle} \\
       c_X \,\text{primal cycle}
}}
\alpha_p^{\,|c_Z^\ast \vee c_X|}
\,
\chi_{e_Z^\ast}(c_Z)\,
\chi_{e_X^\ast}(c_X),
\]
where 
$\alpha_p = 1 - \frac{4p}{3}$,
\(m\) is the number of plaquettes, and \(n\) is the number of qubits.
Here, if \(S\) is a set of edges in the primal (respectively, dual) lattice,
then \(S^\ast\) denotes the corresponding set of edges in the dual
(respectively, primal) lattice.
\end{corollary}

\begin{corollary}\label{depolSurface}
\textnormal{\bf (Kramers--Wannier--MacWilliams duality for planar and rotated surface codes under depolarizing noise).}
 Consider the  planar surface code or the rotated surface code  as defined in Sections~\ref{planarS} and \ref{rotatedS}, with
\(\mathcal{X}\) and \(\mathcal{A}\)  
denoting the underlying primal cell complex  and its 
1-dimensional
subcomplex, \(\mathcal{X}^\ast\) and \(\mathcal{A}^\ast\) denoting their duals, 
and
\(C_X\) and \(C_Z\) the classical codes underlying the CSS construction of these codes. 

Let \(e_Z\) and \(e_X\) be primal and dual errors, represented respectively as
a relative primal \(1\)-chain and a relative dual \(1\)-chain.
Let \(\pi_p(e_Z,e_X)\) denote the probability of the coset 
$(e_Z,e_X) + C_Z^\perp \times C_X^\perp$ 
under the depolarizing channel with parameter \(p\).
Then
\[
\pi_p(e_Z,e_X)
=
2^{-(2n-m-m^\ast)}
\sum_{\substack{
       c_Z \,\text{relative dual cycle} \\
       c_X \,\text{relative primal cycle}
}}
\alpha_p^{\,|c_Z^\ast \vee c_X|}
\,
\chi_{e_Z^\ast}(c_Z)\,
\chi_{e_X^\ast}(c_X),
\] 
where 
$\alpha_p = 1 - \frac{4p}{3}$, 
\(m\) is the number of plaquettes of \(\mathcal{X}\),
\(m^\ast\) is the number of plaquettes of \(\mathcal{X}^\ast\),
and \(n\) is the number of qubits. 
Here, if \(S\) is a set of edges in
\(\mathcal{X}\setminus \mathcal{A}\) (respectively,
\(\mathcal{X}^\ast\setminus \mathcal{A}^\ast\)),
then \(S^\ast\) denotes the corresponding set of edges in
\(\mathcal{X}^\ast\setminus \mathcal{A}^\ast\) (respectively,
\(\mathcal{X}\setminus \mathcal{A}\)).
\end{corollary}

The correlation between \(X\)- and \(Z\)-errors is captured by the bitwise OR in
the term \(\alpha_p^{\,|c_Z^\ast \vee c_X|}\).
In the independent \(X\)- and \(Z\)-noise model, this term is of the form
\(\alpha_p^{\,|c_Z^\ast| + |c_X|}\),
so the sum decomposes into two independent summations, each of which can be
evaluated efficiently.

\begin{openproblem}\label{open2}
Are the joint cycle summations in Corollaries~\ref{depolToric} and 
\ref{depolSurface} efficiently computable?
\end{openproblem}
  Using Fisher gadgets as in Section~\ref{FKT-SMLC}, the joint cycle sum can be
converted into a {\em jointly weighted} sum over all perfect matchings on the
primal decorated graph and all perfect matching  on the dual decorated
graph. However, it is not clear how to efficiently handle the resulting weight
coupling.

\subsection{Proof of Lemma~\ref{genlemDep}} \label{genlemDepProof}
The probability of  \((x,x') \in \mathbb{F}_2^{2n}\) is
\[
(1-p)^{n-|x\vee x'|}
\left(\frac{p}{3}\right)^{|x\vee x'|}
=
(1-p)^n g_r(x,x'),
\]
where
\[
r = \frac{p/3}{1-p}
\quad\text{and}\quad
g_r(x,x') = r^{|x\vee x'|}.
\]
Therefore,
\[
\pi_p(e,e')
=
(1-p)^n
\sum_{y \in C^\perp}\sum_{y' \in C'^\perp}
g_r(e+y,e'+y').
\]
Following the argument in Lemma~\ref{genlem}, 
we expand \(g_r\) in the Fourier basis on \(\mathbb{F}_2^{2n}\):
\[
g_r(x,x') =
\sum_{z,z'} \widehat{g_r}(z,z')\,
\chi_z(x)\chi_{z'}(x').
\]
Using the multiplicativity of characters and rearranging sums gives
\[
\pi_p(e,e')
=
(1-p)^n
\sum_{z,z'}
\widehat{g_r}(z,z')\,
\chi_z(e)\chi_{z'}(e')
\sum_{y \in C^\perp} \chi_z(y)
\sum_{y' \in C'^\perp} \chi_{z'}(y').
\]
By Fact~\ref{ffacts2}, only terms with \(z \in C\) and \(z' \in C'\) contribute,
yielding
\begin{equation}\label{pfbas}
\pi_p(e,e')
=
(1-p)^n |C^\perp|\,|C'^\perp|
\sum_{z \in C}\sum_{z' \in C'}
\widehat{g_r}(z,z')\,
\chi_z(e)\chi_{z'}(e').
\end{equation}
It remains to compute the Fourier transform of \(g_r\):
\begin{eqnarray*}
\widehat{g_r}(z,z') &=& \mathbb{E}_{x,x'} r^{|x\vee x'|}\X_{(z,z')}(x,x')\\
&=& \prod_{i=1}^n \frac{1}{4}\left(1+r(-1)^{z_i}+r(-1)^{z_i'}+r(-1)^{z_i+z_i'}\right). 
\end{eqnarray*}
Observe that
$
(-1)^{z_i} + (-1)^{z_i'} + (-1)^{z_i+z_i'}$ 
evaluates to 
$3$ if   $(z_i,z_i')=(0,0)$, and 
$-1$,   otherwise.
Hence
\[
\widehat{g_r}(z,z')
=
\frac{1}{4^n}
(1+3r)^{n-|z\vee z'|}
(1-r)^{|z\vee z'|}
=
\left(\frac{1+3r}{4}\right)^n
\alpha_p^{|z\vee z'|},
\]
where
\[
\alpha_p = \frac{1-r}{1+3r}.
\]
Substituting this expression into~\eqref{pfbas}, we obtain
\[
\pi_p(e,e')
=
\left(\frac{(1-p)(1+3r)}{4}\right)^n
|C^\perp|\,|C'^\perp|
\sum_{z \in C}\sum_{z' \in C'}
\chi_z(e)\chi_{z'}(e')\,
\alpha_p^{|z\vee z'|}.
\]
The claim then follows 
from  the identities
$(1-p)(1+3r)=1$, 
 $\alpha_p = 1-4p/3$, 
 \(|C^\perp|=2^{n}/|C|\), and  \(|C'^\perp|=2^{n}/|C'|\).
 \finito

\section{Extensions and open questions}\label{extAndOpenQuestions}

For completeness, we note that the techniques developed in this paper extend to the following settings.
 \begin{itemize}
\item {\bf Nonidentically distributed channels.} 
All results in this paper, except the degree-$3$-defect resolution theorems
(Theorems~\ref{DJoin}, \ref{DJoinPlanar}, and~\ref{DJoinRotated}),
extend to channels that are not necessarily identically distributed.

Namely, the SMW decoding algorithm extends by using a degree-$3$ Fisher gadget for each defect 
in the construction of the decorated graph
(Section~\ref{RedSec}),
and by weighting each edge $e$ inherited from the code lattice
by the log-likelihood ratio (LLR)
$L_e = \ln\!\left(\frac{1-p_e}{p_e}\right)$,
where $p_e < 1/2$ is the channel crossover probability of the corresponding qubit.
To turn an instance of MWPM into an instance of MWM, we modify the weights as in
Section~\ref{planarSep}, except that $k$ is now chosen to be the maximum LLR
among all edges.
Everything else in Section~\ref{RedSec} extends without any modification.

As for SML decoding in Section~\ref{FKT-SMLC}, we only need to modify the edge
weights by replacing $(1-2p)$ in the expression of the edge weight
$w_e^{(e_Z)}$ with $(1-2p_e)$.
This follows from extending Lemma~\ref{genlem}  to non-uniform binary symmetric channels
by replacing $(1-2p)$ with $(1-2p_i)$ in the expression of $w_i^{(e)}$,
where $p_i$ is the crossover probability of the 
$i$th bit.
Everything else in Section~\ref{FKT-SMLC} extends without any modification.

\item {\bf Planar surface codes with holes.}
Planar surface codes with holes~\cite{Dennis2002,Fowler2012SurfaceCodes} were introduced to encode multiple logical qubits within a single planar lattice.
 The introduction of holes does not affect the local construction
of decorated graphs, which remains planar.
Thus, the SMW results in Section~\ref{planarSep} extend with the same complexity.
The degree-$3$-defect resolution theorems
(Theorems~\ref{DJoinPlanar} and~\ref{DJoinRotated})
extend as long as the local boundary structure of the underlying lattice near a
hole or a boundary is identical to a local structure
near the boundary of the planar or rotated surface code.
The Pfaffian-based SMLC decoder in Section~\ref{FKT-SMLC} likewise extends by
specializing Lemma~\ref{genlem} to these codes.
Introducing holes, however, requires computing the probabilities of additional
cosets.
As a result, the complexity of the Pfaffian-based SMLC decoder increases by a
constant factor that grows exponentially with the number of holes.

\item {\bf Decoding 2D color codes.} 
The recursive planar separator decoder (Section~\ref{adaptation}) can  be
used to speed up the implementation of the MWPM-based decoder for
two-dimensional color codes proposed by Kubica and Delfosse~\cite{DColor}.
Specifically, one can construct decorated graphs for each of the two restricted
lattices defined in~\cite{DColor},
thereby localizing the corresponding MWPM problems. In each of the decorated graphs, 
defect vertices are represented using odd  Fisher gadgets, while
non-defect vertices are replaced by even  Fisher
gadgets.
Applying the recursive planar-separator-based MWPM solver independently to each of
the two restricted lattices yields an overall runtime of
\(O(n^{3/2}\log n)\).

\end{itemize}

We conclude with a list of open questions motivated by specific results of this paper.

\begin{itemize}
\item 
{\bf Determining the Pfaffian sign in the toric code case.}
Can the sign of the Pfaffian of each matrix $A^{(r,s)}$ in
Theorem~\ref{4paff} be efficiently determined in the context of the toric code?
Doing so would improve the running time of the Pfaffian-based SMLC decoder for the
toric code.

\item 
{\bf Degree-3-defect resolution without bipartiteness.}  
Does the degree-$3$-defect resolution theorem (Theorem~\ref{DJoin})
hold for odd lattice size $L$, i.e., if the toroidal  graph $G$ is not bipartite?
The assumption that $G$ is bipartite is first used in the proof of
Theorem~\ref{DJoin} to show that the walk of bad defects does not visit adjacent
vertices (Lemma~\ref{distinctLemma}), and is used later in
Lemma~\ref{diamondDefects} to derive a contradiction.
Dropping this assumption appears to require a more delicate analysis that allows
visited defects to be adjacent.

\item  {\bf Are simple $D$-joins optimal?}
Consider the toric code graph $G$ and an even-cardinality vertex set $D$,  
and assume that $L$ is even.
If $J$ is a $D$-join, then the $J$-degree of each vertex in $D$ is $1$ or $3$, and
the $J$-degree of each vertex in the complement 
$D^c$ is $0$, $2$, or $4$.
The degree-$3$-defect resolution theorem asserts that there exists a minimum size
$D$-join in which no defect in $D$ has $J$-degree $3$. 
Simulations suggest that a stronger statement may hold:
there exists a minimum size $D$-join in which no defect in $D$ has $J$-degree $3$
and no non-defect in $D^c$ has $J$-degree $4$.
Accordingly, we conjecture that
there exists a minimum size {\em simple} $D$-join, where we
call a $D$-join $J$ simple if the $J$-degree of each vertex in $G$ is at
most $2$.
Equivalently, a simple $D$-join is a union of vertex-disjoint paths connecting
pairs of vertices in $D$.

The original motivation for forbidding degree-$3$ defects was to reduce the size
of the decorated graph.
Strengthening this result by showing that restricting to simple $D$-joins does
not lose optimality would not further reduce the size of the decorated graph, but
it may shed light on the structure of optimal solutions and lead to faster SMW
decoding algorithms.

\end{itemize}

\section*{Acknowledgments} 
The author is grateful to Georges Khater and Ali Sobh for  stimulating  discussions on this work and for helpful  feedback on the first draft of the paper.

\addcontentsline{toc}{section}{References}
\bibliographystyle{IEEEtran} 	
\bibliography{refs}

\newpage

\appendix 

\addcontentsline{toc}{section}{Appendix}
\section*{Appendix}

\section{Proof of Theorem \ref{DJoin}: dissolving degree-$3$ defects}\label{AppendixA}

Let \(G = (V,E)\) denote the graph of the lattice (or its dual) of the
\([[2L^2,2,L]]\) toric code, and assume that \(L\) is even and \(L \ge 4\).

The assumption that \(L\) is {\em even} is needed because the toric-code lattice
\(G\) is {\em bipartite} if and only if \(L\) is even.
Equivalently, when \(L\) is even the graph \(G\) contains no odd-length cycles.
This property is used  in
Lemmas~\ref{distinctLemma} and~\ref{diamondDefects}.

Let \(J\) be a minimum size \(D\)-join.

\begin{figure}
  \centering
  \includegraphics[width=1\textwidth]{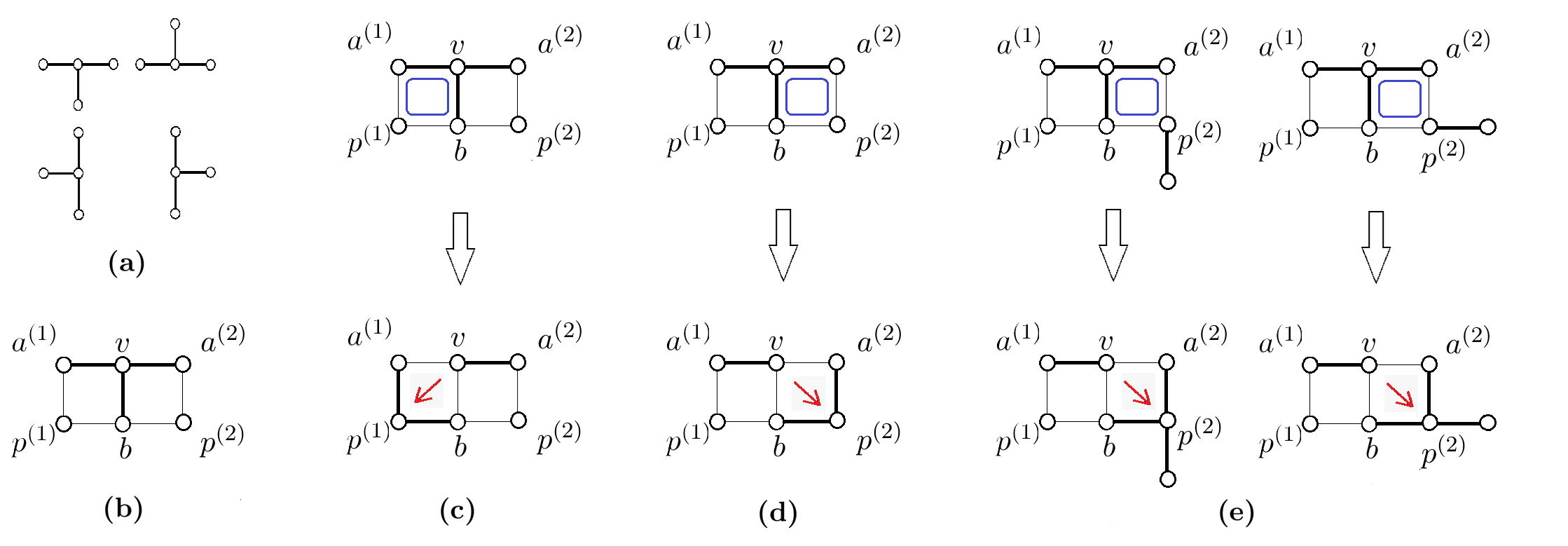}
  \caption{
	{\bf (a)} All possible configurations of \(J\)-edges incident to a \(J\)-bad
defect \(v\).
{\bf (b)} A \(J\)-bad defect \(v\), shown without loss of generality in the first
configuration of Part (a), together with its \(J\)-diagonal partners
\(p^{(1)}\) and \(p^{(2)}\).  Edges belonging to \(J\) are shown in bold.
{\bf (c)} If \(p^{(1)}\) is not a defect, the \(J\)-badness of \(v\) is resolved by
xoring \(J\) with the cycle \(C_{v,p^{(1)}}\), which is shown in blue. 
The upper (lower) panel shows \(J\) before (after) xoring it with the cycle.
{\bf (d)} If \(p^{(2)}\) is not a defect, the \(J\)-badness of \(v\) is resolved by
xoring \(J\) with the cycle \(C_{v,p^{(2)}}\).
{\bf (e)} If \(p^{(1)}\) and \(p^{(2)}\) are defects, the \(J\)-badness of \(v\) is transferred
to \(p^{(2)}\) by xoring \(J\) with the cycle \(C_{v,p^{(2)}}\).
The two columns show the two possible configurations of \(J\)-edges incident to
\(p^{(2)}\).
  }\label{res1}
\end{figure}

\vspace{-0.1in}
\begin{definition}[good/bad defects, diagonal partners]
The elements of \(D\) will be referred to as 
{\em defects}.
A defect is called {\em \(J\)-bad} if its \(J\)-degree is \(3\),
and {\em \(J\)-good} if its \(J\)-degree is \(1\).
Also, define the {\em \(J\)-status} of a defect to be {\em good} or {\em bad},
depending on whether the defect is \(J\)-good or \(J\)-bad.

Consider any \(J\)-bad defect \(v\). We define the  \(J\)-diagonal partners of \(v\) as follows.  
Let \(a^{(1)}\), \(b\), and \(a^{(2)}\) be the three vertices connected to \(v\) by edges
in \(J\), and assume they are labeled so that
\(a^{(1)}\), \(v\), and \(a^{(2)}\) are aligned, as shown in
Figures~\ref{res1}(a) and~\ref{res1}(b).
Let \(p^{(1)}\) be the vertex of \(G\) adjacent to both \(a^{(1)}\) and \(b\),
and let \(p^{(2)}\) be the vertex of \(G\) adjacent to both \(a^{(2)}\) and \(b\).
We call \(p^{(1)}\) and \(p^{(2)}\) the
{\em \(J\)-diagonal partners of \(v\)}. 
Note that \(p^{(1)}\) and \(p^{(2)}\) are {\em distinct} since \(L \neq 2\).  
\end{definition}

\vspace{-0.1in}
To establish Theorem~\ref{DJoin}, we iteratively modify \(J\) by xoring it with
plaquette boundaries, while preserving its size, until no defect is \(J\)-bad. As will be shown below, the defect at \(v\) will either be resolved by a local
xor operation involving one of its \(J\)-diagonal partners, or transferred to
one of them.

Consider the {\em length-$4$ cycle \(C_{v,p^{(1)}}\)}
consisting of the edges of the plaquette containing
\(v\) and \(p^{(1)}\), namely 
\(
C_{v,p^{(1)}} =
\{ (v,a^{(1)}), (a^{(1)},p^{(1)}), (p^{(1)}, b), (b,v) \}.
\) 
Similarly, consider the length-$4$ cycle \(C_{v,p^{(2)}}\) containing
\(v\) and \(p^{(2)}\), namely
\(
C_{v,p^{(2)}} =
\{ (v,a^{(2)}), (a^{(2)},p^{(2)}), (p^{(2)}, b), (b,v) \}.
\)

We consider two cases, depending on whether or not both
\(p^{(1)}\) and \(p^{(2)}\) are defects.

\vspace{-0.1in}
\paragraph{Case 1: At least one of \(p^{(1)}\) or \(p^{(2)}\) is not a defect.}
Suppose that \(p^{(1)}\) is not a defect.
By xoring  \(J\) with
\(C_{v,p^{(1)}}\), we obtain a \(D\)-join \(J'\) of the same size as \(J\),
in which the \(J'\)-degree of \(v\) is equal to \(1\),
while the \(J'\)-degrees of \(a^{(1)}\) and \(a^{(2)}\) coincide with their
\(J\)-degrees.
The degree of \(p^{(1)}\) remains even, and the degrees of all other vertices
of \(G\) are unchanged; see Figure~\ref{res1}(c).

Note that neither \(a^{(1)}\) nor \(b\) can be connected to \(p^{(1)}\) by an
edge of \(J\), since otherwise xoring with \(C_{v,p^{(1)}}\) would strictly
decrease the size of \(J\), contradicting the optimality of \(J\).

Thus, xoring with the length-$4$ cycle \(C_{v,p^{(1)}}\) produces a
\(D\)-join of the same size as \(J\) in which the defect \(v\) becomes
\(J\)-good, while all other defects retain their degrees.
The same argument applies if \(p^{(2)}\) is not a defect, as shown in
Figure~\ref{res1}(d).

Therefore, in Case 1,  the xor operation resolves the issue  that \(v\) is \(J\)-bad.

\vspace{-0.1in}
\paragraph{Case 2: Both \(p^{(1)}\) and \(p^{(2)}\) are defects.}
In this case, both must be \(J\)-good defects; otherwise, \(J\) would not be of
minimum size.
Indeed, if \(p^{(2)}\) were \(J\)-bad, then at least three edges of the cycle
\(C_{v,p^{(2)}}\) would belong to \(J\), since both \(p^{(2)}\) and \(v\) have
\(J\)-degree \(3\).
Xoring \(J\) with \(C_{v,p^{(2)}}\) would then strictly decrease the size of
\(J\), yielding a \(D\)-join smaller than \(J\) and contradicting the optimality
of \(J\).
The same argument applies to \(p^{(1)}\).

By the same reasoning, the unique \(J\)-edge incident to \(p^{(2)}\) does not
belong to the cycle \(C_{v,p^{(2)}}\).
Similarly, the unique \(J\)-edge incident to \(p^{(1)}\) does not belong to the
cycle \(C_{v,p^{(1)}}\).

Thus, xoring \(J\) with \(C_{v,p^{(2)}}\) (or with \(C_{v,p^{(1)}}\)) in Case~2
preserves the size of \(J\) and makes \(v\) \(J\)-good, but renders
\(p^{(2)}\) (or \(p^{(1)}\)) \(J\)-bad, as illustrated in
Figure~\ref{res1}(e).

\bigskip
Therefore, in both cases, the xor operation either resolves the \(J\)-badness of \(v\) or transfers the
 \(J\)-badness of \(v\) to
\(p^{(2)}\) (or \(p^{(1)}\)).

This suggests the following algorithm for resolving bad defects:

\begin{algorithm}
\label{algA}
{\bf Bad-defect resolution algorithm}
\begin{itemize}
\vspace{-3pt}
\item[] Input: \((G,D,J)\)

\item[]
While not all defects are \(J\)-good:
\begin{itemize}
\item[1.] Choose a \(J\)-bad defect \(v\).
\vspace{2pt}
\item[2.] Repeat until \(v\) becomes \(J\)-good:
\begin{itemize}
\vspace{2pt}
\item[3.] Let \(p^{(1)}\) and \(p^{(2)}\) be the \(J\)-diagonal partners of \(v\).
If this is not the first iteration of the inner loop, assume that
\(p^{(1)}\) and \(p^{(2)}\) are ordered so that the previous value of \(v\)
is equal to \(p^{(1)}\).
\vspace{3pt}
\item[4.] If this is the first iteration of the inner loop and
\(p^{(1)}\) is not a defect, update \(J\) by xoring it with the length-$4$
cycle containing \(v\) and \(p^{(1)}\), and break.
\vspace{3pt}
\item[5.] Else if \(p^{(2)}\) is not a defect, update \(J\) by xoring it with
the length-$4$ cycle containing \(v\) and \(p^{(2)}\), and break.
\vspace{3pt}
\item[6.] Else $($i.e., both \(p^{(1)}\) and \(p^{(2)}\) are defects$)$, update \(J\)
by xoring it with the length-$4$ cycle containing \(v\) and \(p^{(2)}\),  and update the value of \(v\) to \(p^{(2)}\).
\end{itemize}
\end{itemize}
\end{itemize}
\end{algorithm}

To show that the algorithm resolves all bad defects, it suffices to show that
it halts. We argue by contradiction.

If the algorithm does not halt, then the inner loop in Line~2 enters an
infinite loop for some \(J\)-bad defect \(v\).
Let \(v_i\) and \(J_i\) denote the values of \(v\) and \(J\), respectively, at
the start of the \(i\)th iteration of the inner loop, for
\(i = 1,2,\ldots\).
Thus, the condition in Line~6 holds for every \(i\).
Hence, for each \(i \ge 1\), \(v_{i+1}\) is a \(J_i\)-diagonal partner of
\(v_i\) and
\(
J_{i+1} = J_i \oplus C_{v_i, v_{i+1}} .
\)

\paragraph{Proof outline}
It is appropriate at this stage to give an overview of the rest of the proof.
We study the evolution of bad defects in
Lemmas~\ref{statusLemma}, \ref{lemmaii}, \ref{distinctLemma},
\ref{lemmaiii}, and~\ref{badDefectEvolution} below.
Together, these lemmas show that
at any step of the algorithm there is a unique
bad defect in the set $S$ of defects visited by the inner loop,
and that this bad defect performs a constrained walk on the toroidal graph.
Using these lemmas, we then show in Lemma~
\ref{badDefectPeriodicStructure} and Corollary~\ref{bipcor}
that if the algorithm does not halt, then the
infinite sequence $v_1, v_2, \ldots$ 
contains all vertices of one of the two bipartitions \(A\) of the toroidal graph \(G\). 
This implies that all vertices in $A$ except one must have
$J$-degree~$1$, with one remaining vertex of $J$-degree~$3$, yielding the
lower bound $|J| \ge L^2/2 + 2$. 
In Lemma~
\ref{diamondDefects} we argue that for any
even set of defects containing $A$, $G$ has a $D$-join $J'$ of size $L^2/2$,
which contradicts the minimality of $J$; hence,
the algorithm must halt.

\begin{lemma}\label{statusLemma}
For all \(i \ge 1\),
\begin{itemize}
\item[a)] The only defects whose \(J_i\)-status differs from their
\(J_{i+1}\)-status are \(v_i\) and \(v_{i+1}\).
In particular, \(v_i\) is \(J_i\)-bad and \(J_{i+1}\)-good, while
\(v_{i+1}\) is \(J_i\)-good and \(J_{i+1}\)-bad.
\item[b)] \(v_i \neq v_{i+2}\).
\end{itemize}
\end{lemma}
{\bf Proof.}
Part~(a) follows from the discussion preceding the algorithm  in Case~2.  
Part~(b) follows from the ordering assumption in Line~3 of the algorithm, and  the update of \(v\) with  
  \(p^{(2)}\) in Line~6.
\finito

Let \(S\) be the set of the \(v_i\)’s without repetitions, i.e.,
\[
S := \{ v_i : i = 1,2,\ldots \}.
\]

\begin{lemma}\label{lemmaii}
For each \(i \ge 1\), \(v_i\) is the unique \(J_i\)-bad defect in \(S\).
\end{lemma} 
{\bf  Proof.}
We argue by induction on \(i\).
The base case is \(i=1\), which we establish by contradiction.
Assume that there exists \(u \neq v_1\) such that \(u\) is \(J_1\)-bad.
Let \(j\) be the smallest index such that \(u = v_j\).
Thus, \(u\) is a \(J_{j-1}\)-good defect.
By the minimality of \(j\) and  Lemma~\ref{statusLemma}(a),
the \(J_k\)-status of \(u\) is the same for 
\(k = j-1,\ldots,1\), which contradicts the assumption that \(u\) is \(J_1\)-bad.
For the induction step, note that the number of \(J_i\)-bad defects in \(S\)
is equal to the number of \(J_{i+1}\)-bad defects in \(S\), 
by Lemma~\ref{statusLemma}(a). 
\finito

Accordingly, for each \(u \neq v_i\), 
\emph{let \(e_i(u)\) denote the unique \(J_i\)-edge incident to
\(u\) at the start of the \(i\)th iteration}.

We investigate below how these edges change as the bad defect is transferred.

The \(i\)th iteration modifies \(J_i\) by xoring it with the length-$4$ cycle
containing \(v_i\) and \(v_{i+1}\).
Thus, the only vertices in the graph whose incident edges are affected are those
lying on this cycle. 
The other two vertices on this cycle, besides \(v_i\) and \(v_{i+1}\), do not belong
to \(S\), by the following lemma which uses the assumption that $L$ is even.

\begin{lemma}\label{distinctLemma}
For each \(i \neq j\), the vertices \(v_i\) and \(v_j\) are not adjacent in \(G\).
\end{lemma}
{\bf Proof.}
For each \(i\), the vertices \(v_i\) and \(v_{i+1}\) are diagonally opposite
vertices of a length-$4$ cycle.
Hence, the distance between \(v_i\) and \(v_{i+1}\) in \(G\) is equal to \(2\).
Therefore, for any \(i \neq j\), the vertices \(v_i\) and \(v_j\) are connected
by a path of even length in \(G\). 
If \(v_i\) were adjacent to \(v_j\), an odd-length cycle would be created in
\(G\), which is impossible since \(G\) is bipartite when \(L\) is even.
\finito

It follows that:

\begin{lemma}\label{lemmaiii}
For all \(i \ge 1\) and all
\(u \in S \setminus \{v_i, v_{i+1}\}\),
we have \(e_{i+1}(u) = e_i(u)\).
\end{lemma}

That is, \(e_i(u)\) remains unchanged as \(v_i\) varies, as long as \(v_i \neq u\).
Whenever \(v_i = u\), the vertex \(u\) becomes a \(J_i\)-bad defect.
The following lemma relates \(e_{i+1}(v_i)\) to \(e_{i-1}(v_i)\) and shows that
the sequence \(v_1, v_2, \ldots\) does not repeat moves in opposite directions.

If \((u,v)\) is an edge of the torus graph \(G\), we define the {\em \(v\)-mirror}
of \((u,v)\) to be the edge \((v,w)\), 
where \(w\) is the unique vertex such that \(u, v, w\) are aligned.

\begin{lemma}\label{badDefectEvolution}
\begin{itemize}
\item[a)]  
  For all \(i \ge 2\), \(e_{i+1}(v_i)\) is the \(v_i\)-mirror of \(e_{i-1}(v_i)\).

\item[b)] 
For all \(i \ge 2\), the edge \(e_{i-1}(v_i)\) does not belong to
\(C_{v_{i-1}, v_i}\) but belongs to \(C_{v_i, v_{i+1}}\).
Its \(v_i\)-mirror \(e_{i+1}(v_i)\) belongs to \(C_{v_{i-1}, v_i}\) but does not
belong to \(C_{v_i, v_{i+1}}\).

Moreover, the cycles \(C_{v_i, v_{i+1}}\) and \(C_{v_{i-1}, v_i}\) share a common
edge, which together with \(e_{i-1}(v_i)\) and its \(v_i\)-mirror
\(e_{i+1}(v_i)\) form the three edges of \(J_i\) incident to \(v_i\).

\item[c)]
For each \(u \in S\), there exist two vertices \(u^+, u^- \in V\), with
\(u^-, u, u^+\) aligned, such that for every \(i \ge 1\) with  \(v_i \neq u\), the edge \(e_i(u)\) is either \((u,u^+)\) or its
\(u\)-mirror \((u,u^-)\).

\item[d)] 
The sequence \(v_1, v_2, \ldots\) does not contain flipped adjacent pairs; that
is,  $(v_{i-1},v_{i})\neq (v_{j+1}, v_{j})$, 
for all $2\leq i\leq j$.

\end{itemize}
  \end{lemma}
{\bf Proof.} 
To establish {\bf Part~(a)}, let \(a,b\) be the two vertices other than
\(v_{i-1}\) and \(v_i\) on the cycle \(C_{v_{i-1},v_i}\).
At the start of iteration \(i-1\), the join \(J_{i-1}\) contains the edges
\((v_{i-1},a)\) and \((v_{i-1},b)\).
At the start of iteration \(i\), these two edges are replaced by
\((a,v_i)\) and \((b,v_i)\) in \(J_i\) by xoring \(J_{i-1}\) with
\(C_{v_{i-1},v_i}\).

\begin{figure}[H]
  \centering
  \includegraphics[width=0.98\textwidth]{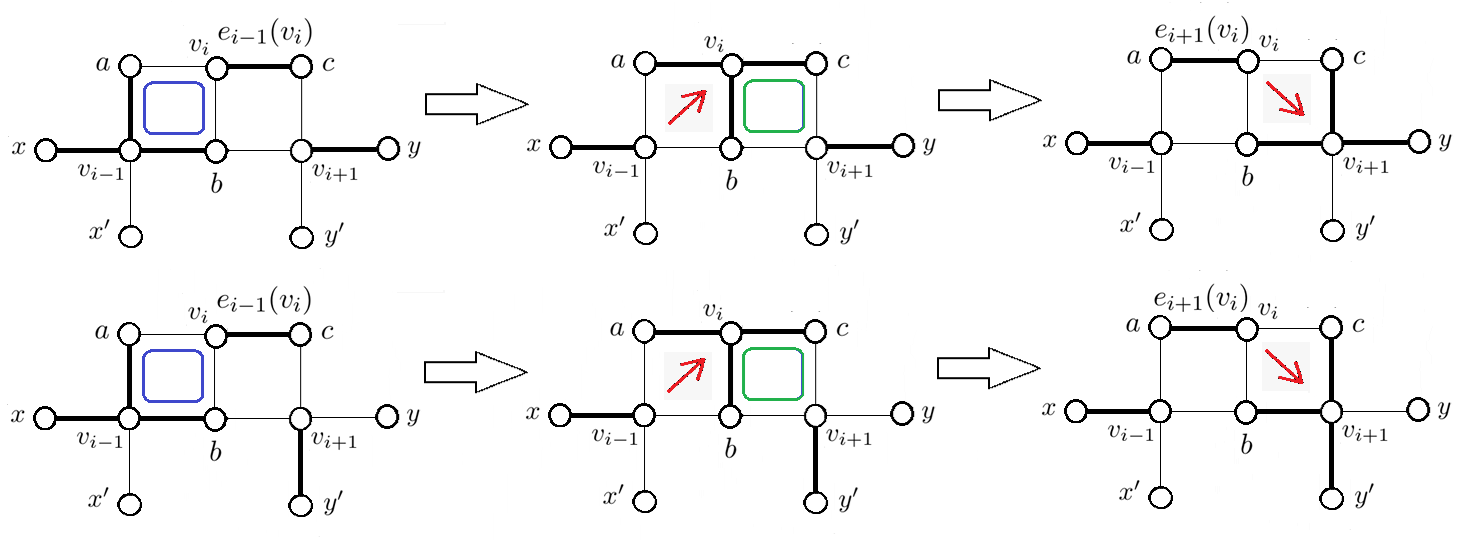}
  \caption{   This figure illustrates the proof of Part~(a) in Lemma~\ref{badDefectEvolution}.
  The neighborhood of \(v_i\) is shown at the start of iterations \(i-1\), \(i\),
  and \(i+1\) in the left, middle, and right panels, respectively.
  The edges of \(J\) are shown in bold, and the cycles
  \(C_{v_{i-1},v_i}\) and \(C_{v_i,v_{i+1}}\) are shown in
  blue and green, respectively.
  The first row shows the case where \((x,v_{i-1}) \in J_{i-1}\) and
  \((v_{i+1},y) \in J_{i+1}\), while the second row shows the case where
  \((x,v_{i-1}) \in J_{i-1}\) and \((v_{i+1},y') \in J_{i+1}\).
  For fixed \(v_i, a, b,\) and \(c\), the remaining two cases---omitted from the figure---correspond to having instead \((x', v_{i-1}) \in J_{i-1}\), which results in a \(45^\circ\) rotation
  of the defect configuration at iteration \(i\).
  }
  \label{res2}
\end{figure}

Since at the start of iteration \(i\), the vertex \(v_i\) is \(J_i\)-bad, it must
 be connected by an edge in  $J_i$ inherited from $J_{i-1}$ to a vertex $c$ 
other than $a$ and $b$, i.e., 
$e_{i-1}(v_i) = (v_i,c)$.  
This vertex \(c\) must be aligned with either \(a, v_i\) or \(b, v_i\).
Without loss of generality, assume that \(a, v_i, c\) are aligned; see
Figure~\ref{res2}.

Since \(v_{i-1} \neq v_{i+1}\) by Lemma~\ref{statusLemma}(b), the
vertices on the cycle \(C_{v_i,v_{i+1}}\) are \(v_i, b, v_{i+1}, c\).
Thus, after xoring with \(C_{v_i,v_{i+1}}\) at the end of iteration \(i\), the
only edge of \(J_{i+1}\) incident to \(v_i\) is \((a,v_i)\).
Hence
\(
e_{i+1}(v_i) = (a,v_i),
\) 
which is the \(v_i\)-mirror of \(e_{i-1}(v_i) = (v_i,c)\).

{\bf Part~(b)} follows directly from the argument in Part~(a).

{\bf Part~(c)} follows by induction from Part~(a) together with Lemma~\ref{lemmaiii}. 
For each \(u \in S \setminus \{v_1\}\), let \(u^{+}\) be the vertex
connected to \(u\) by the edge \(e_1(u)\), and let \(u^{-}\) be the \(u\)-mirror
of \(u^{+}\).
Let \(v_1^{+}\) be the vertex connected to \(v_1\) by \(e_2(v_1)\),
and let \(v_1^{-}\) be the \(v_1\)-mirror of \(v_1^{+}\).
This establishes the base case of the induction. 
Assume that the claim holds for some \(i \ge 1\), and consider iteration
\(i+1\).
By Lemma~\ref{lemmaiii} and the induction hypothesis, the claim holds for all
\(u \in S \setminus \{v_i, v_{i+1}\}\).
By Part~(a), the edge \(e_{i+1}(v_i)\) is the \(v_i\)-mirror of
\(e_{i-1}(v_i)\), and hence the claim also holds for \(v_i\).

{\bf Part~(d)} extends Lemma~\ref{statusLemma}(b).
To establish it, we argue by contradiction.
Assume that there exist indices \(2 \le i \le j\) such that
\((v_{i-1},v_i) = (v_{j+1},v_j)\), and that \(j-i\) is minimal.
We must have \(j>i+1\), since \(j \neq i\) by Lemma~\ref{statusLemma}(b) and
\(j \neq i+1\) because \(v_i \neq v_{i+1}\).

By Part~(b), the cycles \(C_{v_{i-1},v_i}\) and \(C_{v_i,v_{i+1}}\) share an edge,
and each of \(e_{i-1}(v_i)\) and its \(v_i\)-mirror belongs to exactly one of
them.
Similarly, the cycles \(C_{v_{j-1},v_j}\) and \(C_{v_j,v_{j+1}}\) also share an edge,
and each of \(e_{j-1}(v_j)\) and its \(v_j\)-mirror belongs to exactly one of
them.
Since \((v_{i-1},v_i) = (v_{j+1},v_j)\), we have
\(C_{v_{i-1},v_i} = C_{v_j,v_{j+1}}\).
Moreover, by Part~(c), \(e_{i-1}(u)\) is equal to either
\(e_{j-1}(u)\) or its $u$-mirror, where  $u=v_i=v_j$. 
It follows that
\(C_{v_i,v_{i+1}} = C_{v_{j-1},v_j}\), and hence \(v_{i+1} = v_{j-1}\).

Therefore,
\((v_i,v_{i+1}) = (v_j,v_{j-1})\), that is,
\((v_{i'-1},v_{i'}) = (v_{j'+1},v_{j'})\)
for \(i' = i+1\) and \(j' = j-1\),
which contradicts the minimality of \(j-i\), since
\(j' - i' = j - i - 2\). \finito

Since the algorithm is stuck in an infinite loop, there must exist iterations
\(s,t\) such that \(1 \le s < t\) and
\(
(v_s, v_{s+1}) = (v_t, v_{t+1}).
\) 
Assume that \(t-s\) is \emph{minimal} subject to these conditions.

Parts~(a) and~(b) of the following lemma show that both the vertex sequence
\(\{v_i\}_{i \ge s}\) and the edge sequence
\(\{e_i(v_{i+1})\}_{i \ge s}\) are periodic. 
The minimality of $t-s$ is used in the proof of Part~(c).

\begin{lemma}\label{badDefectPeriodicStructure}
 \begin{itemize}
 \item[a)] \(v_{s+k} = v_{t+k}\) for all 
 \(k \ge 0\).
\item[b)] 
 $e_{s+k}(v_{s+k+1}) = e_{t+k}(v_{t+k+1})$ 
for all 
 \(k \ge 0\). 
 \item[c)]  
 In the sequence \(v_s,\ldots, v_{t-1}\), each vertex $v$  appears exactly twice,
and the four vertices of  \(G\) that are diagonally opposite to $v$ 
also appear in the sequence.
Here, two vertices in \(G\) are called {\em diagonally opposite} if they are
non-adjacent vertices of a length-$4$ cycle. 
 \end{itemize}
 \end{lemma}
{\bf Proof.}   The proof  is based on
Lemmas~\ref{statusLemma} and~\ref{badDefectEvolution}. 
To establish {\bf Part~(a)}, we argue by induction on \(k\) that
\(
(v_{s+k}, v_{s+k+1}) = (v_{t+k}, v_{t+k+1}), 
\) 
for all \(k \ge 0\).

The base case \(k=0\) holds by assumption.
Assume that
\((v_{s+k}, v_{s+k+1}) = (v_{t+k}, v_{t+k+1})\).
We show that \(v_{s+k+2} = v_{t+k+2}\).

Let \(v' = v_{s+k} = v_{t+k}\) and
\(v = v_{s+k+1} = v_{t+k+1}\).
By Lemma~\ref{badDefectEvolution}(b),
the edge \(e_{s+k}(v)\) does not belong to the cycle \(C_{v',v}\), but its
\(v\)-mirror does, and the same holds for \(e_{t+k}(v)\).
Hence \(e_{s+k}(v) = e_{t+k}(v)\), and the sets of edges incident to \(v\) in
\(J_{s+k+1}\) and \(J_{t+k+1}\) are identical.
Therefore, the two \(J_{s+k+1}\)-diagonal partners and the two
\(J_{t+k+1}\)-diagonal partners of \(v\) coincide.
The vertex \(v'\) is one such diagonal partner of \(v\) in both cases.
Let \(v''\) be the other one. 
Since \(v_{s+k+2} \neq v_{s+k}\) and \(v_{t+k+2} \neq v_{t+k}\) by
Lemma~\ref{statusLemma}(b), it follows that
\(
v_{s+k+2} = v_{t+k+2} = v''.
\) 
This completes the induction and proves Part~(a).

The same argument also establishes {\bf Part~(b).}

To establish {\bf Part~(c)}, consider any vertex \(v\) appearing in the sequence
\(v_s,\ldots, v_{t-1}\), and let \(i_1\) be the index of its first occurrence.
Assume without loss of generality that
\(i_1 \ge 2\); if \(i_1 = 1\), replace the sequence with the
identical sequence \(v_{t},\ldots, v_{(t-s)+t-1}\).

By Parts~(a) and~(b), we have
\(
e_{i_1-1}(v_{i_1}) = e_{i_1-1+t-s}(v_{i_1}).
\) 
By Lemma~\ref{badDefectEvolution}(a), for each occurrence of \(v\) in the
sequence at iteration $i$, the edge \(e_{i-1}(v)\) is replaced by its \(v\)-mirror at the end
of iteration \(i\).
Hence \(v\) must occur an even number of times in the sequence, and therefore occurs at least twice.

Let $i_1,\ldots, i_m$ be the indices of these occurrences, where
\(2 \le i_1 < \cdots < i_m \le t-1\) and \(m\geq 2\).  
Let \(A_v\) denote the set of four vertices of \(G\) that are diagonally opposite
to \(v\).
For each \(r=1,\ldots,m\), the vertices \(v_{i_r-1}\) and \(v_{i_r+1}\) belong
to \(A_v\).

To prove Part~(c), it suffices to show that
$v_{i_1-1}, v_{i_1+1}, \ldots, v_{i_m-1}, v_{i_m+1}$
are distinct. This implies that $m = 2$ since $|A_v| = 4$.

By Lemma~\ref{statusLemma}(b), we have
\(v_{i_r-1} \neq v_{i_r+1}\) for each \(r\).
Also, $v_{i_r-1}\neq v_{i_{r'}-1}$ for each $r\neq r'$
because if this is not the case, we get
$(v_{i_r-1},v_{i_r}) = (v_{i_{r'}-1},v_{i_{r'}})$,
which contradicts the minimality of $t-s$.
By the same argument, $v_{i_r+1}\neq v_{i_{r'}+1}$ for each $r\neq r'$.
Finally,
$v_{i_r-1}\neq v_{i_{r'}+1}$, for each $r\neq r'$,
because if this is not the case, we get
$(v_{i_r-1},v_{i_r}) = (v_{i_{r'}+1},v_{i_{r'}})$,
which is not possible by Lemma~\ref{badDefectEvolution}(d).
\finito

Consider the set  
$A= \{ v_i : i = s,\ldots, t-1\}$, and recall that
the toroidal graph $G$ is bipartite
since we are assuming that $L$ is even.

\begin{corollary}\label{bipcor}
$A$ is one of the two bipartitions of the toroidal graph $G$.
\end{corollary}
{\bf Proof.}
Since the four diagonally opposite vertices of a vertex in $G$ are those at distance $2$
from this vertex in the toroidal graph, we conclude from
Lemma~\ref{badDefectPeriodicStructure}(c) that
for each vertex $v$ in $A$ and each vertex $u$ at distance
$2$ from $v$ in $G$, $u$ is also in $A$.
By Lemma~\ref{distinctLemma}, $A$ is an independent set.
Thus, it must be the case that $A$ is one of the two bipartitions of  $G$.
\finito

\begin{figure}
  \centering
  \includegraphics[width=0.8\textwidth]{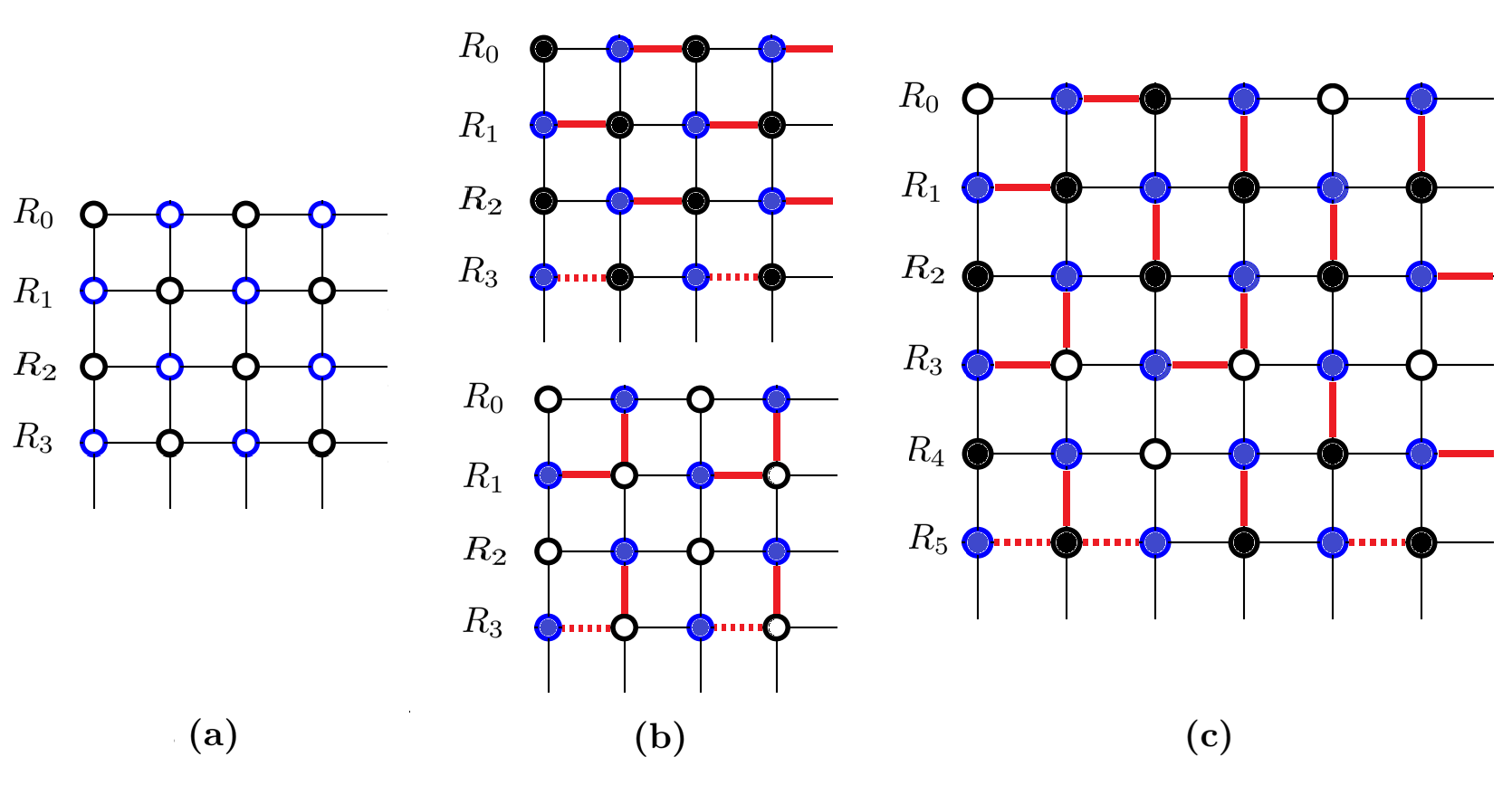}
  \caption{
Panel~{\bf(a)} shows the toroidal graph $G$ for $L=4$, where the first row $R_0$
is identified with the last row $R_4$ and the first column is identified with
the last column. The vertices of the bipartition $A$ are shown in blue.
Panels~{\bf(b)} and~{\bf(c)} illustrate the algorithm used in the proof of
Lemma~\ref{diamondDefects} for $L=4$ and $L=6$, respectively. The vertices of $D$ are filled; the edges of $J$ are shown in red, with those
in $R_{L-1}$ dashed. 
  }
  \label{res3}
\end{figure}

It follows that  $|A| = L^2/2$.  See Figure~\ref{res3}(a) for an illustration of $A$.

Consider the original $D$-join $J = J_1$.
By Lemma~\ref{lemmaii}, each vertex $v$ in $A$ is a $J$-good defect except for
one vertex, which is a $J$-bad defect.
Thus,
\(
|J| \ge 3 + (|A| - 1) = L^2/2 + 2.
\) 
We argue below that this contradicts the minimality of $J$ by 
constructing a $D$-join of smaller size.
Hence Algorithm~\ref{algA} must halt,
which completes the proof of Theorem~\ref{DJoin}.

\begin{lemma}\label{diamondDefects}
Let $A$ be one of the bipartitions of the toroidal graph
$G=(V,E)$, and let
$D$ be an even set of vertices in $G$ containing
$A$.
Then $G$ has a $D$-join $J'$ of size $L^2/2$.
\end{lemma}
{\bf Proof.} 
Consider any ordering $R_0,\ldots, R_{L}$ of the rows of vertices
of $G$ where the first row $R_0$ is identified
with the last one $R_{L}$.
Also, order the vertices in each row according to
the ordering of the columns of $G$, where the first column is identified with
the last.
See Figure~\ref{res4}(a).

Assign a {\em parity} to each vertex in $V$, initialized to
its $D$-{\em parity}, which is defined according to whether or not the vertex
belongs to $D$: {\em odd} parity if it belongs to $D$, and {\em even} parity
otherwise.

Initialize $J'$ to be the empty edge set.
We will populate $J'$ via a greedy algorithm that processes
the vertices one by one.
For each vertex $v$, the algorithm updates $J'$ by including edges incident to $v$.  Let $B$ be the other bipartition  of $G$.
To keep track of changes,
the algorithm updates the parities of the vertices in $B$ without changing
the initial odd $D$-parity of the vertices in $A$, and
while ensuring that once a vertex is processed, its parity is restored to its
original $D$-parity.

Consider the first row $R_0$, and traverse its vertices in order while
considering only those in $A$.
Note that $A \cap R_0$ is one of the bipartitions of the circle graph induced by $G$
on $R_0$. 
For each vertex $v$ in $A$, all four adjacent vertices
are in  $B$.  
We are interested in two of the adjacent vertices to $v$:
the vertex $v_{\text{right}}$ to the right of $v$, and the vertex
$v_{\text{down}}$ in the following row.
If $v_{\text{right}}$ has odd parity, add to $J'$
an edge connecting $v$ and $v_{\text{right}}$.
Otherwise, add to $J'$ an edge connecting $v$ to $v_{\text{down}}$,
and flip the parity of $v_{\text{down}}$.
Thus, after processing the first row, the parity of the $J'$-degree of each
vertex in the first row is still equal to its $D$-parity.
Moreover, the parities of all $B$-vertices in the second row
are properly updated according to the edges included
so far in $J'$.
The total number of edges added to $J'$ is $L/2$.

Repeat the same process on $R_1$, then $R_2$, and so on, and stop right before
reaching the row $R_{L-1}$.
Thus, so far $|J'| = (L-1)L/2$.
At this stage, the parity of the $J'$-degree of each vertex in the first $L-1$
rows is equal to its $D$-parity.
The parity of each $B$-vertex $v$ in $R_{L-1}$ is equal to its
$D$-parity if it
is not connected by a $J'$-edge to a vertex in $R_{L-2}$; otherwise, the parity
of $v$ is equal to the negation of its $D$-parity.

We do not process the last row $R_L$ because it is identified with the first
row.
As for $R_{L-1}$, we process it differently because we do not want to affect
the parity of the vertices already processed in $R_0$.
We show below how to process this row using
$L/2$
horizontal edges.
The key observation is that since the size of $D$ is even, the number $m$ of
vertices in $R_{L-1}$ with odd parity is even.

Let $C_{L-1}$ be the cycle graph 
induced by $G$ 
on the vertices in
$R_{L-1}$,
$A_{L-1} = A \cap R_{L-1}$, and $B_{L-1} = B \cap R_{L-1}$.
Thus, $A_{L-1}$ and $B_{L-1}$ form the bipartition of the cycle graph
$C_{L-1}$, $|A_{L-1}| = |B_{L-1}| = L/2$, and each of
$A_{L-1}$ and $B_{L-1}$
contains every other vertex of $C_{L-1}$.

Let $Q$ be the set of odd-parity vertices in $R_{L-1}$.
Thus, $A_{L-1} \subset Q$ and $m = |Q|$ is even.
Since $m$ is even, $C_{L-1}$ has a $Q$-join.
Consider any such $Q$-join $K$. 
Since $C_{L-1}$ is bipartite, the number of edges in $K$ is equal to the sum of the $K$-degrees of all vertices in $A_{L-1}$.
Since $A_{L-1} \subset Q$, the
$K$-degree of each vertex in $A_{L-1}$ is $1$, and thus
$|K| = L/2$.
The theorem then follows from adding the edges
in $K$ to $J'$.

See Figure~\ref{res3}(b,c) for an illustration of the algorithm.
\finito

\section{Proof of Theorems~\ref{DJoinPlanar} and~\ref{DJoinRotated}: 	planar and rotated surface codes} \label{AppendixB}

We explain in this section how the proof of Theorem~\ref{DJoin} adapts to the planar and rotated
surface codes.
The local bad-defect resolution steps are unchanged in the interior of the lattice.
In the planar surface code case, we show that the walk of bad defects does not reach
the boundary. In the rotated surface code case, we show that the bad defect is resolved if the walk reaches the boundary.
As for the set $A$ whose elements have their diagonally opposite vertices in it, it
does not exist in the case of planar and rotated surface codes because they lack the
periodic structure.
We conclude that the bad-defect resolution algorithm terminates because $A$ does not
exist.
In what follows, we elaborate on the adaptation by considering the cases of the
planar and rotated surface codes separately.

\subsection{Proof of Theorem~\ref{DJoinPlanar}}
Consider the planar surface code as defined in Section~\ref{planarS}, in terms of the 
$L\times (L+1)$ lattice $\mathcal{X}$ and its $1$-dimensional sublattice $\mathcal{A}$. 
Let $B$ be the set of vertices of $\mathcal{A}$, and let $G = (V,E)$ denote the subgraph obtained from 
$\mathcal{X}$ by excluding the edges in $\mathcal{A}$. 
Assume that $L \geq 3$.

Let $D \subset V$ be a subset of vertices of $G$ disjoint from $B$, and let $J$ be a minimum size $D$-join $J$ relative to $B$. To turn $J$ into a $B$-relative $D$-join  in which
all defects in $D$ are $J$-good, we will apply the bad-defect resolution algorithm,
Algorithm~\ref{algA}, without any modification.

\begin{figure}
  \centering
  \includegraphics[width=0.9\textwidth]{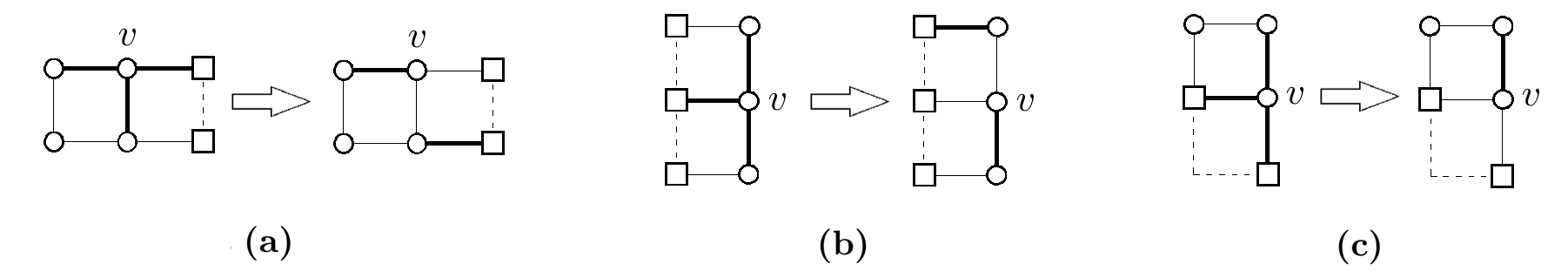}
  \caption{ 
  Panels {\bf (a)} and {\bf (b)} show all possible cases, up to rotation and mirroring,
of a $J$-bad defect
that has a $J$-diagonal partner in $B$,
with respect to a $B$-relative $D$-join $J$
in the case of the {\bf planar surface code}.
The left subpanel shows the edges of $J$ in bold.
The vertices in $B$ are indicated by boxes, 
and the dashed edges are edges in $\mathcal{A}$ (or $\mathcal{A}^\ast$). 
The right subpanel shows that the edges of $J$ can be updated
while still forming a $B$-relative $D$-join, but with fewer edges.
Panel {\bf (a)} shows the case when only 
one of the $J$-diagonal partners of $v$ is in $B$.
Panel {\bf (b)} shows the case when both $J$-diagonal partners of $v$
are in $B$. 
Panel {\bf (c)} shows all possible cases, up to rotation and mirroring,
of a $J$-bad defect
that has a $J$-diagonal partner outside $G'$,
with respect to a $B$-relative $D$-join $J$
in the case of the {\bf rotated surface code}. 
The dashed edges are edges in $\mathcal{A}$ (or $\mathcal{A}^\ast$), and the missing diagonal partner is the vertex
incident to both dashed edges.
The left subpanel shows the edges of $J$ in bold, with the vertices in
$B$ indicated by boxes.
The right subpanel shows that the edges of $J$ can be updated
while still forming a $B$-relative $D$-join, but with fewer edges.
  }
  \label{res4}
\end{figure}

To see why this algorithm is applicable to the new setup, it is enough to note
that the $J$-diagonal partners
$p^{(1)}$ and $p^{(2)}$ of a defect $v$ in Algorithm~\ref{algA} cannot be  in $B$.
The case when only one of them belongs to $B$ is not possible,
as shown in Case~(a) of Figure~\ref{res4},
and the case when both of them are in $B$ is not possible either,
as shown in Case~(b) of Figure~\ref{res4}.  
 Both cases contradict the minimality of $J$.

To show that the algorithm halts, we need to slightly adjust the argument:
\begin{itemize}
\item 
The proof of Lemma~\ref{distinctLemma}
relies on the bipartiteness of $G$.
In the setup of Theorem~\ref{DJoin}, the toroidal graph $G$ is bipartite because $L$ is even.
In the setup of Theorem~\ref{DJoinPlanar}, $G$ is bipartite because it is a subgraph
of a grid whose cells are squares, and in any such grid all cycles are of even length.
Thus, the statement of Lemma~\ref{distinctLemma} holds.

\item 
The set $A = \{ v_i : i = s,\ldots, t-1 \}$ does not exist in
$G$ because $G$ does not have a periodicity that
allows for such a structure.
The fact that each $v$ in $A$ has four diagonally opposite vertices in $A$
would make $A$ infinite.

\item Therefore 
Lemma~\ref{diamondDefects} is not applicable, and 
the nonexistence of $A$ yields the desired contradiction.
\end{itemize}
The same argument applies to the dual lattice $\mathcal{X}^\ast$ and
its one-dimensional sublattice $\mathcal{A}^\ast$.

\subsection{Proof of Theorem~\ref{DJoinRotated}}

Consider now the $[[L^2,1,L]]$ rotated surface code and let $G$ be the graph of
$\mathcal{X}$ (or $\mathcal{X}^\ast$) obtained by excluding the edges of
$\mathcal{A}$ (or $\mathcal{X}^\ast$) and deleting the vertices of
$\mathcal{A}$ (or $\mathcal{A}^\ast$) that become isolated, and let $B$ be the
remaining set of vertices of $\mathcal{A}$ (or $\mathcal{A}^\ast$).

Before running the bad-defect resolution algorithm, we delete from $G$ the
degree-$1$ vertices in $B$. 
The graph $G$ has $2$ or $4$ such vertices, unless it is associated with
the dual lattice and $L$ is even (see Figure~\ref{rotated}).
If $J$ has an edge $(u,v)$ incident to a degree-$1$ vertex $v$ in $B$,
we update $J$ by removing $(u,v)$ from $J$ and then
xoring $J$ with the other edge connecting $u$ to a degree-$2$ vertex in $B$.
This keeps $J$ a $B$-relative $D$-join and does not increase the size of $J$.
Hence the size must remain the same, since $J$ is minimal.

We then run the bad-defect resolution algorithm on the resulting graph $G'$.
The reason we need to remove these vertices is that, due to the diamond structure
of the graph $G$, the $J$-diagonal partners of a $J$-bad defect $v$ may not
belong to the graph $G$ if $v$ is near the boundary.
Removing them ensures that the diagonal partners must be present in the graph,
since $J$ has minimal size, as illustrated in Case~(c) of Figure~\ref{res3}.

In the setup of the planar surface code, we argued that
the $J$-diagonal partners $p^{(1)}$ and $p^{(2)}$
of a bad defect $v$ in Algorithm~\ref{algA}
cannot be in $B$.
This does not hold for the primal or dual diamond lattices of the rotated surface code, but it does not matter  since the vertices in $B$ are not defects.
That is, if $p^{(1)}$ is in $B$, then
this must be the first iteration and the inner loop breaks at Line~4.
If $p^{(2)}$ is in $B$, then
the inner loop breaks at Line~5.

Theorem~\ref{DJoinRotated} follows from  adapting  the analysis of the
bad-defect resolution algorithm, as  in the proof of
Theorem~\ref{DJoinPlanar}.
Lemma~\ref{distinctLemma} holds because here again $G'$
is a subgraph of a grid whose cells are squares,
hence all cycles are of even length.
The set $A$ does not exist, hence
Lemma~\ref{diamondDefects} is not applicable, and
the nonexistence of $A$ yields the desired contradiction.

\section{Pfaffians and Pfaffian orientations}
\label{app:pfaffians}

See~\cite{LovaszPlummer} for a general reference on matchings, Pfaffians, and
Pfaffian orientations. In what follows, we briefly recall the relevant preliminaries needed in
this paper.

Let $G=(V,E)$ be an \emph{undirected graph} on an even-cardinality vertex set
$V=\{1,\ldots,2k\}$, and let $\mathcal{M}_G$ denote the set of all
\emph{perfect matchings} of $G$.

For each $M\in\mathcal{M}_G$, consider an \emph{orientation} $\vec M$ of the edges
in $M$; that is, for each edge $(i,j)\in M$, exactly one of $(i,j)$ or $(j,i)$
belongs to $\vec M$.
Fix also an arbitrary \emph{ordering} $(i_1,j_1),\ldots,(i_k,j_k)$ 
of the oriented edges in $\vec M$.
Define the permutation $\pi_{\vec M}\in S_{2k}$ mapping
$(1,\ldots,2k)$ to $(i_1,j_1,\ldots,i_k,j_k)$.
The \emph{sign of the oriented matching} $\vec M$,
denoted $\operatorname{sgn}(\vec M)$, is defined as the sign of this permutation.
The sign depends on the orientation of the edges in $M$, but not on the ordering
of the edges.

\medskip
Let $A$ be a real $2k \times 2k$ \emph{skew-symmetric matrix}, i.e.,
$A_{ij} = -A_{ji}$ for all $i,j$, supported on the edges of $G$; that is,
$(i,j) \in E$ if $A_{ij} \neq 0$.
The \emph{Pfaffian} of $A$ is defined by
\begin{equation}\label{pafexp}
\operatorname{Pf}(A)
=
\sum_{M\in\mathcal{M}_G}
\operatorname{sgn}(\vec M)
\prod_{(i,j)\in \vec M} A_{ij}.
\end{equation}
Although  $\operatorname{sgn}(\vec M)$ depends on the orientation $\vec M$, 
the term $\operatorname{sgn}(\vec M)
\prod_{(i,j)\in \vec M} A_{ij}$ 
 does not depend 
on this orientation since $A$ is
skew-symmetric.

In 1842, Cayley showed that $\operatorname{Pf}(A)^2=\det(A)$ 
for every skew-symmetric matrix $A$.

\medskip

Consider now a \emph{global orientation} $\vec E$ of the edges of $G$; that is,
for each edge $(i,j)\in E$, exactly one of $(i,j)$ or $(j,i)$ belongs to $\vec E$.
For a perfect matching $M$, let $\vec M$ denote the orientation of the edges of
$M$ induced by $\vec E$.

The orientation $\vec E$ is called a \emph{Pfaffian orientation} of $G$ if 
$\operatorname{sgn}(\vec M)=\operatorname{sgn}(\vec M')$, 
for all perfect matchings $M,M'\in\mathcal{M}_G$.
Thus, under a Pfaffian orientation all perfect matchings contribute with
the same sign to the Pfaffian expansion in (\ref{pafexp}).

\medskip

Consequently, if $G=(V,E)$ is a weighted undirected graph admitting a Pfaffian
orientation $\vec E$, with edge weights $w:E\to\mathbb{R}$, 
consider the weighted sum over perfect matchings
\[
S=\sum_{M\in\mathcal{M}_G}~\prod_{e\in M} w_e.
\]
Then
\[
S=\pm\operatorname{Pf}(A),
\]
where the skew-symmetric matrix $A$ is defined by
\[
A_{ij}=
\begin{cases}
\phantom{-}w_{ij}, & (i,j)\in\vec E,\\
-w_{ij}, & (j,i)\in\vec E,\\
0, & \text{otherwise}.
\end{cases}
\]

\medskip

In a seminal work, Kasteleyn \cite{Kasteleyn1963} showed that every  
 2-connected 
planar graph admits a Pfaffian
orientation, which can be constructed efficiently. 
As a consequence, the absolute value of the weighted sum over perfect matchings of a planar
graph can be computed in polynomial time by evaluating the determinant of the
associated skew-symmetric matrix and taking its square root.

\end{document}